\documentclass[review]{elsarticle}

\journal{Journal of \LaTeX\ Templates}

\usepackage{graphicx}
\usepackage{booktabs} 
\usepackage[utf8]{inputenc} 
\usepackage[T1]{fontenc}    
\usepackage{hyperref}       
\usepackage{url}            
\usepackage{booktabs}       
\usepackage{amsfonts}       
\usepackage{nicefrac}       
\usepackage{microtype}      
\usepackage{algorithm,algorithmicx,algpseudocode,amsmath}
\usepackage{epstopdf}
\usepackage{bbm}
\usepackage{subfigure}

\newtheorem{myDef}{Definition}
\newtheorem{myTheo}{Theorem}
\newtheorem{myLemma}{Lemma}	
\newtheorem{myProof}{Proof}		

\begin{document}

\begin{frontmatter}

\title{FLDP: Flexible strategy for local differential privacy}

\author{Dan Zhao}
\author{Suyun Zhao}
\author{Ruixuan Liu}
\author{Cuiping Li}
\author{Wenjuan Liang}

\author{Hong Chen\corref{mycorrespondingauthor}}
\cortext[mycorrespondingauthor]{Corresponding author}
\ead{chong@ruc.edu.cn}
\ead[url]{teacher.zjut.cc/t34881/}

\address{Key Laboratory of Data Engineering and Knowledge Engineering of Ministry of Education (RenminUniver-sity), Beijing 100872
	
	School of Information, Renmin University of China, Beijing 100872}

\begin{abstract}
Local differential privacy (LDP), a technique applying unbiased statistical estimations instead of real data, is often adopted in data collection. In particular, this technique is used with frequency oracles (FO) because it can protect each user's privacy and prevent leakage of sensitive information. However, the definition of LDP is so conservative that it requires all inputs to be indistinguishable after perturbation. Indeed, LDP protects each value; however, it is rarely used in practical scenarios owing to its cost in terms of accuracy. In this paper, we address the challenge of providing weakened but flexible protection where each value only needs to be indistinguishable from part of the domain after perturbation. First, we present this weakened but flexible LDP (FLDP) notion. We then prove the association with LDP and DP. Second, we design an FHR approach for the common FO issue while satisfying FLDP. The proposed approach balances communication cost, computational complexity, and estimation accuracy. Finally, experimental results using practical and synthetic datasets verify the effectiveness and efficiency of our approach.
\end{abstract}


\end{frontmatter}


\section{Introduction}
\label{intro}

In data collection, which is the foundation of data mining and analysis, there are several threats of invading users privacy and leaking sensitive information. Moreover, the disclosure of individual information may create crises of trust and financial losses, among other negative outcomes. Differential privacy as the \emph{de facto} standard for private data release was first introduced by Dwork\cite{dwork2006calibrating}. 

In recent years, local differential privacy (LDP) has been proposed to avoid the requirement of a central trusted authority. This mechanism uses the concept of differential privacy in the data collection stage, which means that the perturbed mechanism exists on the local side. LDP applies unbiased statistical estimations instead of real data to protect each user's privacy and prevent the leakage of sensitive information. Moreover, LDP has been widely adopted in industry, for example, companies such as Google\cite{erlingsson2014rappor}, Apple\cite{team2017learning}, and Microsoft\cite{ding2017collecting}.

Under the notion of LDP, each data owner encodes their values and sends this information to an aggregator after perturbation: An adversary cannot distinguish any pair of values in the domain with high confidence (this is controlled by a privacy budget $\varepsilon$). However, LDP is overly conservative in that it requires each input to be indistinguishable from any other input after perturbation, which requires substantial noise. For example, the classical mechanism RAPPOR hashes each value into $h$ bits on a vector of length $d$. To ensure that the outputs of any two different values are similar, each bit of the vector must be disturbed under the uniform privacy budget. From another perspective, only $h$ bits are valid, and the remaining $d-h$ bits are added noise required to satisfy LDP. Indeed, excessive noise does better protect each value, but practical scenarios can rarely use this notion due to the accuracy reduction it entails.

\textbf{Motivation.}
Based on this motivation, we define a weak but flexible version of LDP (FLDP) that does not need all the outputs to have the same range after perturbation but requires intersection of the output.

To satisfy LDP, the input should be hidden over the entire domain; however, to satisfy FLDP, any input should be hidden in part or all of the domain. Although FLDP weakens privacy, it makes mechanism design more flexible in addressing different practical application scenarios. Under the condition that weak privacy protection is required, the intersection of different ranges could decrease rather than increase the privacy budget, which may lead to easy leakage of the real value. Under conditions of strong privacy protection, the intersection can be enlarged to include the entire universe (satisfying the original LDP). That is, FLDP relaxes LDP such that the perturbation mechanism does not require overly strong privacy protection, where LDP would overprotect the inputs that are less sensitive.

In this study, we design an effective notion called FHR for a frequency oracle (FO) on single-item data while satisfying FLDP. From a practical perspective, we consider three aspects:  privacy protection, query accuracy, and communication and calculation. This paper then introduces and analyzes the FHR mechanism in detail and presents results of its application in an experiment.

The main contributions of this work are summarized as follows:
\begin{itemize}
	\item We introduce a new privacy notion called FLDP, which allows a more flexible design mechanism than that of LDP for different real scenarios. 
	\item We design an notion called FHR for a FO with an unbiased estimator that satisfies FLDP. We introduce and analyze the proposed FHR mechanism in detail.
	
	\item We validate the flexibility and effectiveness of our notion and mechanism and empirically demonstrate the effectiveness of the FHR mechanism using synthetic and practical datasets.
\end{itemize}

\textbf{Roadmap} In Section 2, we discuss the related work. We then present background for the development of our notion in Section 3. In Section 4, we propose our notions of LDP. We then discuss the design a mechanism that satisfies our notion in Section 5. Section 6 describes the methods used to assess the proposed approach. The experimental results are presented in Section 7. Finally, Section 8 concludes this paper.

\section{Related Work}

Differential privacy (DP) as the \emph{de facto} standard of data privacy was first introduced by Dworkd\cite{dwork2006calibrating}. Centralized DP has attracted considerable attention, including theoretical treatments \cite{dwork2014algorithmic,vadhan2017complexity} and practical perspectives \cite{li2016differential,papernot2018scalable}. In a local setting, without a trust aggregator, all users hide their input via perturbation mechanisms. The local privacy model was first formalized in 2011 \cite{kasiviswanathan2011can}. Google\cite{erlingsson2014rappor} proposed RAPPOR for frequency estimation and applied it for the first time to collect homepage data in the Chrome browser. Subsequently, various methods such as O-RR, O-RAPPOR\cite{kairouz2016discrete}, and RAPPOR-unknown\cite{fanti2016building} were developed as modifications of RAPPOR for specific scenarios. Moreover, LDP techniques have been applied to FOs\cite{bassily2015local,wang2017locally} and mean values\cite{duchi2013local,bassily2015local}.
Consequently, LDP has been implemented by several companies, such as Apple\cite{team2017learning}, Samsung\cite{SamsungDP}, and Microsoft\cite{ding2017collecting}.

However, LDP is not as widely used as DP because DP requires little noise, whereas LDP requires substantial noise. Thus, several variants of LDP and their corresponding mechanisms have been studied. For example, personalized LDP (PLDP) \cite{chen2016private} divides users into different groups according to different privacy budgets. In contrast, condensed LDP\cite{gursoy2019secure}, utility-optimized LDP\cite{murakami2019utility}, and input-discriminative LDP (IN-LDP)\cite{gu2020providing} group the inputs based on defined rules to improve transmission accuracy. The purpose of these definitions is to weaken the LDP notion and make it less conservative, as there is no real need for such noise in practical scenarios. This paper proposes a privacy notion called FLDP, which are more flexible than LDP.

\section{Preliminary Knowledge}

\subsection{LDP Notion}
LDP is a local model of DP for collecting user data without a credible aggregator
\cite{bassily2017practical,bun2018heavy,qin2016heavy,wang2017locallyprivate,wang2017locally}. An LDP mechanism $\mathcal{M}$ ensures that the probability of one value being sent to the server approximates the probability of any other values being sent. The formal privacy requirement satisfies $\varepsilon$-LDP as follows:
\begin{myDef}[$\varepsilon$-Local Differnetial Privacy]\label{LDP def}  
	Given a mechanism $\mathcal{M}$ with domain $\mathcal{X}$ and range $\mathcal{R}$, if any two items $t$ and $t'$($t,t' \in \mathcal{X}$) output the same value $s$($s \in \mathcal{R}$) through mechanism $\mathcal{M}$, which satisfies the inequality \ref{LDP mechanism}, then we say that $\mathcal{M}$ satisfies $\varepsilon$-LDP\cite{dwork2008differential}.
	
	\begin{equation}
	\centering
	\label{LDP mechanism}
	Pr[\mathcal{M}(t) = s] \le e^\varepsilon\cdot Pr[\mathcal{M}(t') = s]
	\end{equation}
\end{myDef}

We can adjust the privacy budget $\varepsilon$ to balance data availability and privacy. Moreover, LDP can provide a stronger level of privacy protection than that of a centralized setting because each user reports only the perturbed data.

\subsection{Problem Statement}
\textbf{System Model.} Our system model involves one aggregator and $n$ users $\mathcal{U}=\{u_1,u_2,...,u_n\}$. Each user possesses one item $t$ in a finite universe $\mathcal{I}=\{1,2,...,d\}$ and adds random noise independently before sending information to the aggregator. Then, the aggregator collects users' data and learns statistical information based on all users while protecting the privacy of each individual user. 
However, the LDP notion is conservative. We assume that any two items $t_i \in I$ and $t_j\in I$ have ranges $\mathcal{R}(t_i)$ and $\mathcal{R}(t_j)$ via perturbation mechanism $\mathcal{M}$. If $\mathcal{M}$ satisfies LDP, then $\mathcal{R}(t_i) = \mathcal{R}(t_j)$. 

Thus, each item $t$ is hidden in the domain $\mathcal{I}$. In a practical application scenario, such a strict requirement is not needed: It is sufficient for item $t$ to be hidden in part or all of the domain $\mathcal{I}$. This is a weaker but more flexible privacy strategy than LDP that also satisfies centralized DP on the server side. In particular, a FO represents common statistical information and is the core issue in LDP research. An FO represents the protocols enabling estimation of the frequency of any value in the domain $\mathcal{I}$ \cite{cormode2018privacy}.

\textbf{Threat Model.} 
We assume that each user is honest and not malicious, but that different users or a user and the aggregator can collude and that transmission can be monitored.  Therefore, the most stringent attacker has access to all users’ transmitted data and to the perturbation mechanism and its parameters, except for the target user.

\textbf{Utility Goals.} 
The utility goals of the mechanisms are that the attacker cannot infer the true user’s value from the disturbance information sent by the user while still allowing the aggregator to obtain an estimated statistic. This enables the availability of user data while protecting the privacy of each user. Thus, the objective of the designed mechanism is to obtain unbiased estimates and low variances.

\subsection{Frequency Oracle protocols}
Frequency Oracle is a core issue under the LDP framework, which has attracted a lot of theoretical and practical attention. The protocols that are enabling to estimate the frequency of any item/itemset in the domain $I$ are called Frequency Oracle (FO). We review the state-of-the-art LDP protocols on FO. Our mechanism is inspired by previous approaches but also contrasts with them. We want to define a more flexible LDP, and we can design algorithms to achieve more accurate FO.

\subsubsection{Random Response (RR)\cite{warner1965randomized}.}
A classic RR technique commonly used in statistics can be adapted to LDP \cite{qin2016heavy}. RR requires the user to return a random response to protect sensitive information. In particular, each user provides a true answer with probability $p$ or gives a random answer with probability $1-p$, where $p=\frac{e^\varepsilon}{e^\varepsilon +1}$, to satisfy $\varepsilon$-LDP.  Thus, we can substitute an approximation $\hat{f}$ instead of the real frequency using the formula $\hat{f} = \frac{f}{(2p-1)} + \frac{(p-1)}{(2p-1)}$, where $f$ is the collected frequency. However, the standard RR is limited to binary data.

\subsubsection{Generalized Random Response(GRR)\cite{wang2018locally}.}
GRR is used to address $d$ classification. All users send their own values with probability $p$ or randomly send  $d-1$ other values with the remaining probability. The perturbation function is formally defined as

\begin{equation}
\label{Generalized Random Response}
P[\mathcal{M}_{GRR}(t)=s]=\left\{
\begin{aligned}
p= \frac{e^{\varepsilon}}{e^{\varepsilon}+d-1}, \qquad \textrm{if}\ s=t\\
q= \frac{1}{e^{\varepsilon}+d-1}, \qquad \textrm{if}\ s\neq t
\end{aligned}
\right.
\end{equation}
\subsubsection{RAPPOR\cite{erlingsson2014rappor} and Optimized Unary Encoding (OUE)\cite{wang2017locallyprivate}.}
RAPPOR and OUE encode items into a vector and send the vector to the aggregator after perturbation. Next, the aggregator implements summation to obtain the total count of each bit in the vector, denoted by $c_i$ for the $i$-th bit. Therefore, the aggregator obtains an unbiased estimator via $\hat{c_i}=\frac{c_i-nq}{p-q}$. Moreover, OUE can provide higher utility than RAPPOR for frequency estimation under the same $\varepsilon$ owing to its use of optimization.

\subsubsection{Optimized Local Hashing (OLH)\cite{wang2017locallyprivate}.}
OLH addresses a large domain size $d$ by first using a random hash function to map an item into a smaller domain of size $g$ (the optimal choice is $\lceil \varepsilon+1 \rceil$ ) and then applying GRR to the hash value in the smaller domain, where $p=0.5$ and $q=\frac 1g$. In OLH, the reporting protocol $<H^j,y^j>$ for the user $u_j$ is 
$$
M_{OLH}(t):= <H,M_{GRR}(H(v))>
$$
where $H$ is randomly chosen from a family of hash functions.

The aggregator computes the number of reports via $C(t)=|{j|H^j(t)=y^j}|$ for each item $t\in I$. Then, the aggregator can obtain the unbiased estimate via $\hat{f_t}=\frac{c(t)-n/g}{p-1/g}$.


\begin{table}[t]
	\caption{Notation}
	\label{NodationDefinitions}
	\centering
	\begin{tabular}{ll}
		\toprule
		\cmidrule{0-1}
		Notation & Definition   \\
		\midrule
		$\varepsilon$   & privacy budget\\
		$n$				& total number of users\\
		$\mathcal{I}$ 	& domain of users' items\\
		$\mathcal{R}(t_i)$ & domain of the item $t_i$ via perturbation\\
		$t_i$			& value that user $u_i$ holds\\
		$H_r$			& Hadamard matrices of order $2^r$ for every nonnegative integer $r$\\
		$b_i$			& vector sent via perturbation mechanism\\
		$\hat{z}$		& summation and correction of $\hat{z}$\\
		\bottomrule
	\end{tabular}
\end{table}

\section{Flexible LDP}

This section introduces a new privacy notion called FLDP, which provides weaker protection but is more flexible than LDP.

\subsection{Definition}
LDP requires any input to be indistinguishable from any other input after perturbation. However, LDP is too conservative to use in practical applications because the definition is based on the worst-case scenario. Intuitively, we need any item in the partial domain $\mathcal{I}$ to be indistinguishable after a perturbation mechanism.

%
%

\begin{myDef}[$(\varepsilon,\eta)$-FLDP]\label{F-LDP def}  
	Given a mechanism $\mathcal{M}$ with domain $\mathcal{I}$ and range $\mathcal{R}$, if any items $t,t'$ have the range $R(t),R(t')$ via mechanism $\mathcal{M}$ respectively. If the range satisfies the inequality \ref{equ: eta} and the output satisfies the inequality \ref{eau: pr}, then we say that $\mathcal{M}$ satisfies $(\varepsilon,\eta)$-FLDP.
		\begin{equation}
	\centering
	\label{equ: eta}
	\min_{t,t'\in \mathcal{I}}\frac{|R(t)\cap R(t')|}{\max\{|R(t)|,|R(t')|\}} \geq \eta 
	\end{equation}
	\begin{equation}
	\centering
	\label{eau: pr}
	\max_{s\in R(t)\cap R(t')} \frac{Pr[\mathcal{M}(t) = s]}{Pr[\mathcal{M}(t') = s] }\le e^\varepsilon 
	\end{equation}
	
\end{myDef}

In Definition \ref{F-LDP def}, any output $s$ can be transformed by the element of a group $\mathcal{G}\subseteq \mathcal{I}$ via perturbation mechanism $\mathcal{M}$; thus, any element $t\in \mathcal{G}$ can be hidden in the group $\mathcal{G}$. We can control the privacy protection intensity not only via the perturbation parameter $\varepsilon$ but also by changing the size of the group $\mathcal{G}$. Therefore, the easiest way to satisfy FLDP is to divide the input into groups, which can be done using the current LDP protocols. We assume that the size of the group is $m$; then the number of groups is $d/m$. If the size of the groups in mechanism $\mathcal{M}$ satisfies $m=d$, which means that $\mathcal{G}=\mathcal{I}$, then $\mathcal{M}$ satisfies $\varepsilon$-LDP. 
 
This notion is equivalent to collecting the statistic for each group separately; thus, we define a stronger FLDP to protect privacy.

%
%

\subsection{Relationship with Other Notions}
As a mature technique, central DP has been used widely in industry. The DP notion protects the privacy of overall statistical information, but it requires a trusted aggregator. Thus, partially participating users do not trust the aggregator. Moreover, local DP provides worst-case privacy protection for all users and all inputs using more noise. This conservative strategy improves the overall strength of the resulting privacy but decreases its applicability. This paper proposes the notion of FLDP, which is an intermediate between DP and LDP. The FLDP notion not only guarantees the privacy of statistics (Lemma \ref{lemma: satisfy DP}), but also protects the local data to some extent (Lemma \ref{lemma: satisfy FLDP}). Therefore, FLDP is flexible and has extensive possible applications.

\begin{myLemma}
	\label{lemma: satisfy FLDP}
	$\varepsilon$-LDP is a special definition of ($\varepsilon,\eta$=1)-FLDP.
\end{myLemma}
\begin{myProof}
	If $\eta$=1,  $R(t)=R(t')=R(t)\cap R(t')$. It means that if an output $s$ is from $t$, then $Pr[\mathcal{M}(t') = s]\neq 0$. The domains of output are the same between the definition  of $\varepsilon$-LDP and ($\varepsilon,\eta$=1)-FLDP.
%
\end{myProof}
\begin{myLemma}
	\label{lemma: satisfy DP}
	FLDP approximately satisfies DP. 
\end{myLemma}
\begin{myProof}
	By means of the FLDP mechanism, we can obtain an unbiased estimate with variance $\sigma^2$.
	Since FLDP protocols are based on LDP, the estimated frequency conforms to a binomial distribution. According to the central limit theorem, the estimate $\hat{f_t}$ can be viewed as the true value $f_t$ plus normally distributed noise. 
	\begin{equation}
	\hat{f_t} \approx f_t + \mathcal{N}(0,\sigma^2)
	\end{equation}
	Then, we deem that FLDP approximately satisfies the central DP.
\end{myProof}

\begin{figure}[htp]
	\centering
	\begin{center}
		\graphicspath{{Img/}}
		\includegraphics[width=0.95\textwidth]
		{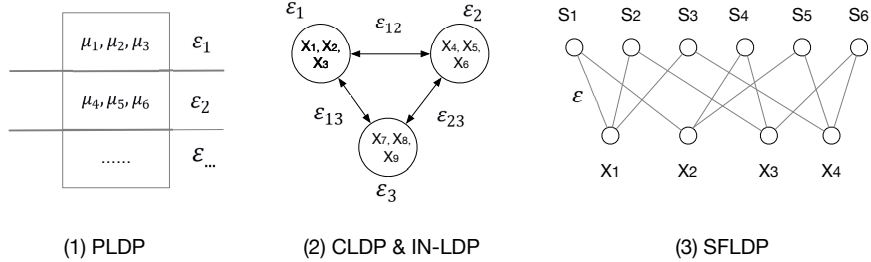} %
	\end{center}
	\caption{Different LDP notions, where $u$ stands for user, $x$ for input, and $s$ for output.}
	\label{Fig: LDP_notions}
\end{figure}

There exist some other notions that are loosely considered as versions of LDP, such as PLDP\cite{yiwen2018utility}, CLDP\cite{gursoy2019secure}, and ID-LDP\cite{gu2020providing}, shown in Figure \ref{Fig: LDP_notions}. In brief, the classification of the privacy budget in the PLDP notion is based on different user requirements: to provide discriminative privacy for inputs, CLDP and IDLDP set the privacy budget for each possible pair based on defined rules. If viewed from this perspective, FLDP also sets rules on the input. Thus, CLDP and In-LDP are special cases of FLDP. FLDP sets the transition probability of two inputs with weak correlation to zero and sets the conversion parameter of any two inputs in a group to $\varepsilon$. This mechanism improves the efficiency and accuracy of each transmission.

\section{Perturbation Mechanism for Frequency Oracle under FLDP}

In this section, we address the practical issue at hand: application to  FO. To address this challenge, we design an algorithm Flexible Hadamard Response (FHR) to obtain an unbiased frequency estimator while satisfying ($\varepsilon,\eta$=0.5)-FLDP. Then we analyze this perturbation mechanism for the FO under FLDP in detail.

\subsection{Mechanism Design}
Our goal is to design a framework with perturbation mechanism $\mathcal{M}$ and a FO protocol that satisfies the proposed FLDP notion. The challenge of design mechanism $\mathcal{M}$ is that the ranges $\mathcal{R}(t)$ and $\mathcal{R}(t')$ feature subtraction and intersection for any two different inputs $t$ and $t'$ ($\mathcal{R}(t)\cap \mathcal{R}(t')\neq \emptyset$, $\mathcal{R}(t)\verb|\| \mathcal{R}(t')\neq \emptyset  $ and $\mathcal{R}(t')\verb|\| \mathcal{R}(t)\neq \emptyset  $). Furthermore, to reduce communication and computation, we propose the FHR notion by introducing the Hadamard matrix to encode the input. Algorithm \ref{FHR algorithm} presents the pseudocode for FHR, which is divided into three phases: \emph{encoding}, \emph{perturbation}, and \emph{frequency estimation}.

\begin{algorithm}[htb]
	\caption{Algorithm: Flexible Hadamard Response(FHR).}
	\label{FHR algorithm}
	\begin{algorithmic}[1]
		\Require
		The privacy budget $\varepsilon$; Hadamard matrix $H$.
		\Ensure
		The estimated frequency $f_t$;
		\For{each user $u_i$ with value $t_i$, $i=1$ to $n$ }
		
		\State Obtainthe article readable the vector $H(t_i)$ in $H$ corresponding to item $t_i$;
		\State Randomly select one $+1$ and one $-1$ from $H(t)$ to obtain the vector $b_i^*$;
		\State Sample a Bernoulli variable $u$ that equals 1 with probability $p=\frac{e^\varepsilon}{e^\varepsilon+1}$;
		\If {$u=1$ }
		\State $b_i$ = $b_i^*$
		\Else
		\State $b_i$ = $-b_i^*$
		\EndIf
		\State Send $b_i$ to the server;
		\EndFor
		\State The server corrects the summation $\hat{z}\leftarrow \frac{(e^{\varepsilon}+1)}{2(e^{\varepsilon}-1)}\cdot(\sum\limits_{i=1}^{n} b_{i})$;
		\State Obtain the FO for any item $t$  with $\hat{f_t}=\hat{b}\leftarrow \hat{z}\cdot H(t)$;\\
		\Return $\hat{f_t}$;
	\end{algorithmic}
\end{algorithm}

\textbf{Encoding.} (step 2)
We first encode each item into a $2^r$-length vector mapping from Hadamard matrix $H_r$, except for the first row, where $r=\lceil2^{log_2{(|\mathcal{I}|+1)}} \rceil $  (unless otherwise stated, all the symbols $H$ below represent $H_r$).  Thus, user $u_i$ with item $t_i$ can obtain a vector $H(t_i)$ via Hadamard matrix mapping. The Hadamard matrix is generated by the formula \ref{generation of hadamard matrix}.
\begin{equation}
\begin{aligned}
\label{generation of hadamard matrix}
H_{r+1}=\left[\begin{matrix}
H_{r} & H_{r}\\
H_{r} & -H_{r}
\end{matrix}\right]
\end{aligned}
\end{equation}
The property of the encoding is that half of the encoded vector values are $+1$ and the others are $-1$. In addition, the positions of a vector at $+1$ or $-1$ have half values $+1$ and half values $-1$ in any other vector.

\textbf{Perturbation.} (steps 3$\sim$11)
In this phase, we randomly select one $+1$ value  and one $-1$ value  from the encoding vector, which are denoted as $(x,y)$. We then use a random response to perturb this pair. Each user holds the true values $(x,y)$ with probability $p$ and the reverse values $(-x,-y)$ with probability $1-p$. In theory, user $u_i$ has a vector $b_i$ of length $2^r$ that has values only at the positions of $x$ and $y$ and has value 0 at all other positions. Thus, each user sends $(index_x,x,index_y)$ to the server, and the communication cost is $2r+1$. Because $y$ is the inverse of $x$, we do not need to send the value of $y$. Moreover, on the user side, the value of the corresponding position can be rapidly obtained based on the row and column of the Hadamard matrix. We can obtain the value in the $i$-th row and $j$-th column $[i, j]$ by $H[row,col]=(-1)^{Count\mathbbm{1}(bin(i\&j))}$, which means that we first conduct the binary operation between $i$ and $j$ and then count the number of 1s.

\textbf{Frequency Estimation.} (steps 12$\sim$14)
The aggregator obtains all the vectors sent by the users and an unbiased frequency estimate is obtained by the calculation. First, the aggregator adds vectors sent by the user to obtain a summation vector $\hat{z}=\sum_i^n b_i$ of length $2^r$. Thereafter, the corresponding Hadamard vector $H(t)$ of any item $t$ can be calculated with $\hat{z}$ to obtain the corresponding frequency $\hat{f_t}$ (\ref{frequency estimation}).
\begin{equation}
\label{frequency estimation}
\hat{f_t}=\frac{(e^{\varepsilon}+1)}{2(e^{\varepsilon}-1)}\cdot(\hat{z}\cdot H(t))
\end{equation}

\subsection{Privacy and Analysis}

\subsubsection{Privacy Guarantee}
Two theorems establish the privacy and accuracy
guarantee of FHR: Theorem \ref{theo: FLDP} proves that FHR satisfies ($\varepsilon,\eta$=0.5)-FLDP while Theorem \ref{theo:unbiased} proves that the frequency estimation is unbiased. These two theorems guarantee the privacy and availability of FHR.

\begin{myTheo}  
	\label{theo:unbiased}
	The correction $\hat{f_t}$ is unbiased. 
	\begin{myProof}		
		We divide users into $n_t$ and $n-n_t$ for item $t$. If $b_i$ is generated from item $t$, then $b_i \cdot H(t)$ corresponds to the $\mathcal{X}$ distribution \ref{X distribution}. Clearly, each participant gives 2 with probability $p$ or $-2$ with probability $1-p$. 
		
		If $b_i$ is generated from any other item $t'\in I \verb|\| \{t\}$, then $b_i \cdot H(t)$ corresponds to the $\mathcal{Y}$ distribution \ref{Y distribution}, which has four cases of equal probability values ($p_1=p_2=p_3=p_4=0.25$): the positions of values equal to two are $+1$ in $H(t)$; the positions of values equal to two are $-1$ in $H(t)$; the position of the positive value is $+1$ in $H(t)$ and that of the negative value is $-1$ in $H(t)$; the position of the positive value is $-1$ in $H(t)$ and that of the negative value is $+1$ in $H(t)$. In the first and second cases, $b_i \cdot H(t)$ is always equal to 0. 
		\begin{equation}
		\label{X distribution}
		\mathcal{X}=\left\{
		\begin{aligned}
		2 \qquad& p=\frac {e^{\varepsilon}}{e^{\varepsilon}+1}\\
		-2 \qquad& 1-p=\frac {1}{e^{\varepsilon}+1}
		\end{aligned}
		\right.
		\end{equation}
		\begin{equation}
		\label{Y distribution}
		\mathcal{Y}=\left\{
		\begin{aligned}
		0 \qquad& p_1+p_2=0.5\\
		2 \qquad& p_3\times p+p_4\times (1-p)=0.25\\
		-2 \qquad& p_4\times p+p_3\times (1-p)=0.25
		\end{aligned}
		\right.
		\end{equation}		
		Thus, we prove this Theorem as follows:
		$$
		\begin{aligned}
		\label{unbiased f_t}
		\mathbb{E}(\hat{f_t})
		&=\frac{(e^{\varepsilon}+1)}{2(e^{\varepsilon}-1)}\mathbb{E}\sum\limits_{i=1}^{n}(b_i \cdot H(t)) \\
		&=\frac{(e^{\varepsilon}+1)}{2(e^{\varepsilon}-1)}\bigg(\mathbb{E}\sum\limits_{i=1}^{n_t}(b_i \cdot H(t))+\mathbb{E}\sum\limits_{i=1}^{n-n_t}(b_i \cdot H(t))\bigg)\\
		&=\frac{(e^{\varepsilon}+1)}{2(e^{\varepsilon}-1)}\bigg(\sum\limits_{i=1}^{n_t}\mathbb{E}\mathcal{X}+\sum\limits_{i=1}^{n-n_t}\mathbb{E}\mathcal{Y}\bigg)=n_t
		\end{aligned}
		$$
	\end{myProof}
\end{myTheo}

\begin{myTheo}
	\label{theo: FLDP}
	Algorithm FHR satisfies ($\varepsilon,\eta$=0.5)-FLDP.
	\begin{myProof}
		Suppose there are two users $u_1$ and $u_2$ with values $t_1$ and $t_2$. Then, the ranges are $R(t)$ and $R(t')$. Let $d$ denote the length of the vector. There are a total of $|R(t)|=|\mathcal{R}(t')|=\frac {d^2}4$ outputs for each item via mechanism $\mathcal{M}$, and the intersection between the two ranges is $|\mathcal{R}(t)\cap \mathcal{R}(t')|=\frac {d^2}8$. For any output $s\in \mathcal{R}(t)\cap \mathcal{R}(t')$, the sensitivity is maximized when $\mathcal{M}_{t}(x,y)$ and $\mathcal{M}_{t'}(x,y)$ are opposites.
		$$
		\begin{aligned}
		RC_{max}
		&=\max{\frac {P(s|t)}{P(s|t')}} \\
		&=\frac {P((x,y)_s=(1,-1)|(x,y)_t=(1,-1))}{P((x,y)_s=(1,-1)|(x,y)_{t'}=(-1,1))} \\
		&=e^\varepsilon
		\end{aligned}
		$$	
The output $s\in \mathcal{R}(t)\verb|\| \mathcal{R}(t')$ or $\mathcal{R}(t')\verb|\| \mathcal{R}(t) $, we can calculate $\eta = \frac {|\mathcal{R}(t)\cap \mathcal{R}(t')|}{|R(t)|}=0.5$.
Then FHR satisfies ($\varepsilon,\eta$=0.5)-FLDP.
		
	\end{myProof}
\end{myTheo}

\subsubsection{Accuracy Analysis}
We analyze the accuracy improvement of the proposed framework
by evaluating the error bound in Theorem \ref{theo:unbiased}, which is independent of the value of the perturbation mechanism. We first calculate the variance of the estimated frequency.  The upper bound noise of each item is then $O(\frac{\sqrt{\log{(1/\beta)}}}{\varepsilon \sqrt{n}})$ in Lemma \ref{Lmmma: union bound}. Finally, we compare the variance obtained with FHR with that obtained using other LDP mechanisms.

\textbf{Variance.}
The variance in the frequency of item $t$ in FHR is denoted as follows:
$\hat{f_t}=\frac{(e^{\varepsilon}+1)}{2(e^{\varepsilon}-1)}\cdot(\hat{z}\cdot H(t))$
\begin{equation}
\label{variance FHR}
\begin{aligned}
\mathcal{D}(f_t)
&=\frac{(e^{\varepsilon}+1)^2}{4(e^{\varepsilon}-1)^2}\sum\limits_i^n(b_i\cdot H(t))\\
&=\frac{(e^{\varepsilon}+1)^2}{4(e^{\varepsilon}-1)^2}\bigg(\sum\limits_i^{nt}\mathcal{D}(\mathcal{X})+\sum\limits_i^{n-nt}\mathcal{D}(\mathcal{Y}) \bigg)\\
&=\frac{(e^{\varepsilon}+1)^2}{2(e^{\varepsilon}-1)^2}n+\bigg(\frac {(e^{\varepsilon}+1)^2}{2(e^{\varepsilon}-1)^2} -1 \bigg)n_t\\
&\le \frac{(e^{\varepsilon}+1)^2}{2(e^{\varepsilon}-1)^2}n
\end{aligned}
\end{equation}

\begin{myLemma}[Upper bound]
	\label{Lmmma: union bound}
	Let $d=2^r$. As described earlier, $f_t=\frac 1d\sum_{i=1}^n(H(t)\cdot H(t))$ and $\hat{f_t}= \frac{e^{\varepsilon}+1}{2(e^{\varepsilon}-1)}\sum_{i=1}^n(b_i\cdot H(t)) $. With at least $1-\beta$ probability,
	$$
	\max|f_t-\hat{f_t}|=O(\frac{\sqrt{\log{(1/\beta)}}}{\varepsilon \sqrt{n}})
	$$
\end{myLemma}
\begin{myProof}
	In the FHR algorithm, $\hat{f_t}$ is the unbiased counterpart of $f_t$ obtained by \ref{theo:unbiased}. Let $w_i=\frac 1d(H(t)\cdot H(t))$ and $\hat{w_i}=\frac{e^{\varepsilon}+1}{2(e^\varepsilon-1)}(b_i\cdot H(t))$. Thus, $|w_i-\hat{w_i}|\leq 2$. Then, according to Bernstein's inequality,
	\begin{equation}
	\label{Bernsterin FHR}
	\begin{aligned}
	&Pr[|f_t-\hat{f_t}|>\lambda]
	=Pr[|\sum_{i=1}^n(w_i-\hat{w_i})|>n\lambda]\\
	&\leq 2\cdot\exp{\Bigg(-\frac {n^2\lambda^2}{2\sum_{i=1}^nVar(\hat{w_i})+\frac 43n\lambda }\Bigg)}
	\end{aligned}
	\end{equation}
	
	The variance is obtained by formula \ref{variance FHR}  $\sum_{i=1}^nVar(\hat{w_i})=O(\frac n{\varepsilon^2})$.	Thus, we obtain
	$$
	\begin{aligned}
	&Pr[|f_t-\hat{f_t}|>\lambda]
	&\leq 2\cdot\exp{\Big(-\frac {n^2\lambda^2}{O(n/\varepsilon^2)+n\lambda O(1)}\Big)}
	\end{aligned}
	$$
	By the union bound, there exists 
	\begin{equation}
	\label{ana:lamdba}
	\lambda=O(\frac{\sqrt{\log{(1/\beta)}}}{\varepsilon \sqrt{n}})
	\end{equation}
	such that $\max|f_t-\hat{f_t}|< \lambda$ holds with a probability of at least $1-\beta$. 
\end{myProof} 

For OLH and OUE, the optimal variance is $\frac{4\varepsilon}{(\varepsilon+1)^2}$. According to formula \ref{variance FHR}, FHR performs best when $\varepsilon<ln(sqrt(8)+3)\approx 1.76$.  Generally, the privacy protection effect is lost when the privacy parameter $\varepsilon$ is greater than 2. Thus, only cases where $\varepsilon$ is larger than 2 are discussed in academia, and the actual available localization privacy budget is usually less than $2$. 

\subsubsection{Complexity}

We analyzed complexity in terms of user communication and aggregator computation. Table \ref{tab:complexity} compares various FO mechanisms with FHR. Since each user sends two locations and one value, the communication cost is $O(\log{|\mathcal{I}|})$; this may not be the lowest possible cost in the real world but is still very small. Considering the calculation complexity in the aggregator, FHR requires one dot product of the matrix in order to obtain the FO. The operation efficiency is very high, and FHR has the shortest run time in the experiment compared with other methods.

\begin{table}
	\caption{Complexity of FO mechanisms}
	\label{tab:complexity}      
	\begin{tabular}{|c|c|c|c|c|}
		\hline\noalign{\smallskip}
		& OLH & RAPPOR & OUE & FHR  \\
		\noalign{\smallskip}\hline\noalign{\smallskip}
		Communication & $\log{n}+\log(g)$ & $|\mathcal{I}|$& $|\mathcal{I}|$ & $2\log{|\mathcal{I}|}+1$ \\ \hline 
		Calculation & $\varOmega(n|\mathcal{I}|)$ & $\varOmega(|\mathcal{I}|)$ & $\varOmega(|\mathcal{I}|)$ & $\varOmega(|\mathcal{I}|)$ \\
		\noalign{\smallskip}\hline
	\end{tabular}
\end{table}

Overall, from the perspective of practical application, the privacy budget of FHR has the smallest variance within a reasonable range, the communication required is minimal, and the computational complexity is the lowest compared with alternative approaches. Thus, FHR is proposed for use in real situations.

\section{Experiment}
\subsection{Material and Competitors}

\textbf{Environment.}
All mechanisms are implemented in Python 3.7.3, and all the experiments are conducted on an Intel Core(TM) i7-6700 3.40 GHz PC with 16 GB memory. We obtained and reported average results over 10 runs.

\textbf{Datasets.}
The following datasets were considered for the experiments. 
\begin{itemize}
	\item Zipf: We generate a \emph{Zipf} dataset with 593358 records and $1023$ different values. The Zipf dataset is generated randomly according to Zipf's law because the Zipf distribution is similar to the real distribution, and it is easy to simulate the influence of the parameters in different cases. We generate this dataset using the \emph{numpy.random.zipf()} function. 
	\item  Online:  This dataset contains the $2363344$ merchant records with 2603 categories from online retail.
	\item NLTK:  We use the dataset from the NLTK BROWN corpus \cite{Nltkdata}. There are $1008320$ records with $49585$ different unique words.
\end{itemize}

\textbf{Metrics.}
We evaluate the data availability in terms of the distributional difference and accuracy. We adopt the Kullback--Leibler divergence (KLD) \cite{kairouz2014extremal}, related error \cite{li2012privbasis}, squared error \cite{wang2018locally}, and normalized cumulative rank (NCR) \cite{wang2017locally,wang2018locally} to assess the difference and accuracy of these mechanisms.

a). Kullback--Leibler divergence. $Csisz \acute{a} rs$ \emph{f-divergence} is used to measure whether a privacy mechanism is an information-theoretic quantity, and the distributional difference is defined in formula \ref{oldKLD} \cite{kairouz2014extremal}.
\begin{equation}
\label{oldKLD}
D_f\big(R_{est}||R_{real}\big) = \int f\bigg(\frac{dR_{real}}{dR_{est}}\bigg)dR_{est}
\end{equation}
where $f(x) = x \log{x}$. To express divergence more accurately, formula \ref{NewKLD} is adopted in this study to detect data availability.
\begin{equation}
\label{NewKLD}
KLD =\frac 12 \Big(D_f\big(R_{est}||R_{real}\big)+D_f\big(R_{real}||R_{est}\big)\Big)
\end{equation}

b). Related error (RE)\cite{li2012privbasis}. RE is another metric for the reliability of the estimated frequencies. For any candidate item $c\in C$, let $p_c$ and $p^*_c$ denote the real and estimated frequency, respectively:
\begin{equation}
RE= median_{c\in C} \frac{|p_c-p^*_c|}{p_c}
\end{equation}

c). Squared error (SE)\cite{wang2018locally}. SE is also a metric used to evaluate estimation frequency. That is, 
\begin{equation}
SE= \frac{1}{|\textbf{X}_{est}\cap \textbf{X}_{real}|} \sum_{c\in  \textbf{X}_{est}\cap \textbf{X}_{real}}(p_c-p^*_c)^2
\end{equation}

d). Normalized cumulative rank (NCR) \cite{wang2017locally,wang2018locally}.
The quality of a function is ranked according to the $k$ value as follows: The highest ranked value has a score of $k$, the next one has a score of $k-1$, and so on.

The $k$-th value has a score of 1, and all other values have scores of 0. To normalize this into a value between 0 and 1, we divide the sum of scores by the maximum possible score, i.e., $\frac{k(k+1)} 2$. 

\textbf{Competitors and parameter setting.}  Since OLH and OUE are the most common and efficient mechanisms used in FO problems, we compare them with FHR to demonstrate their effectiveness. Meanwhile, this study considers practical application as the premise and a small privacy budget setting of $0.4\leq\varepsilon\leq 2$.

\subsection{Results and Discussion}

This section discusses the evaluation of FO based on the results obtained.

\begin{figure*}[!htp]
	\centering
	\graphicspath{{Img/}}
	\subfigure[ZipfData for Top 20]{
		\includegraphics[width=0.29\textwidth]{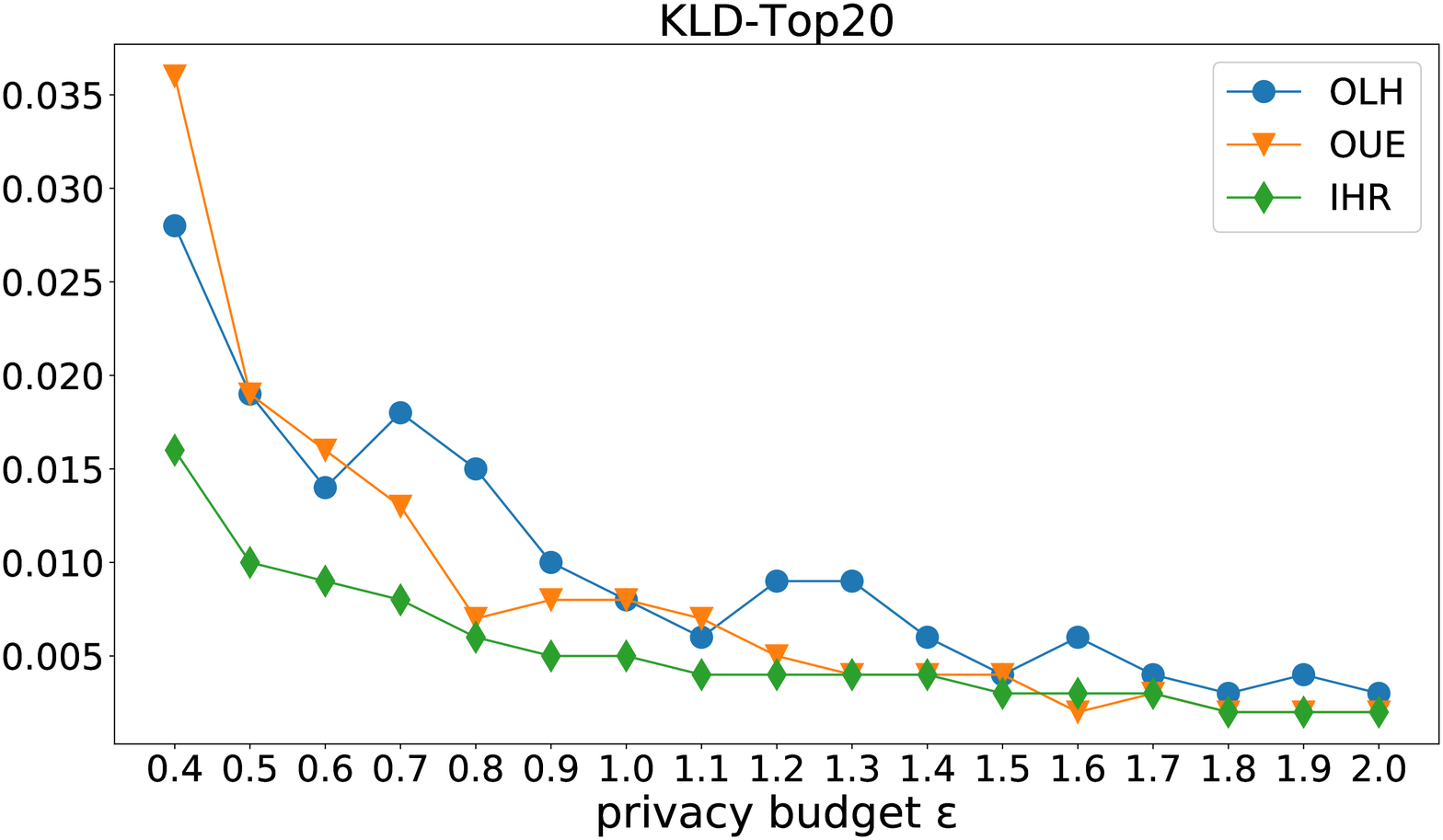}
		\label{fig:ZipfData_Top20_KLD}}
	\subfigure[Online for Top 20]{
		\includegraphics[width=0.29\textwidth]{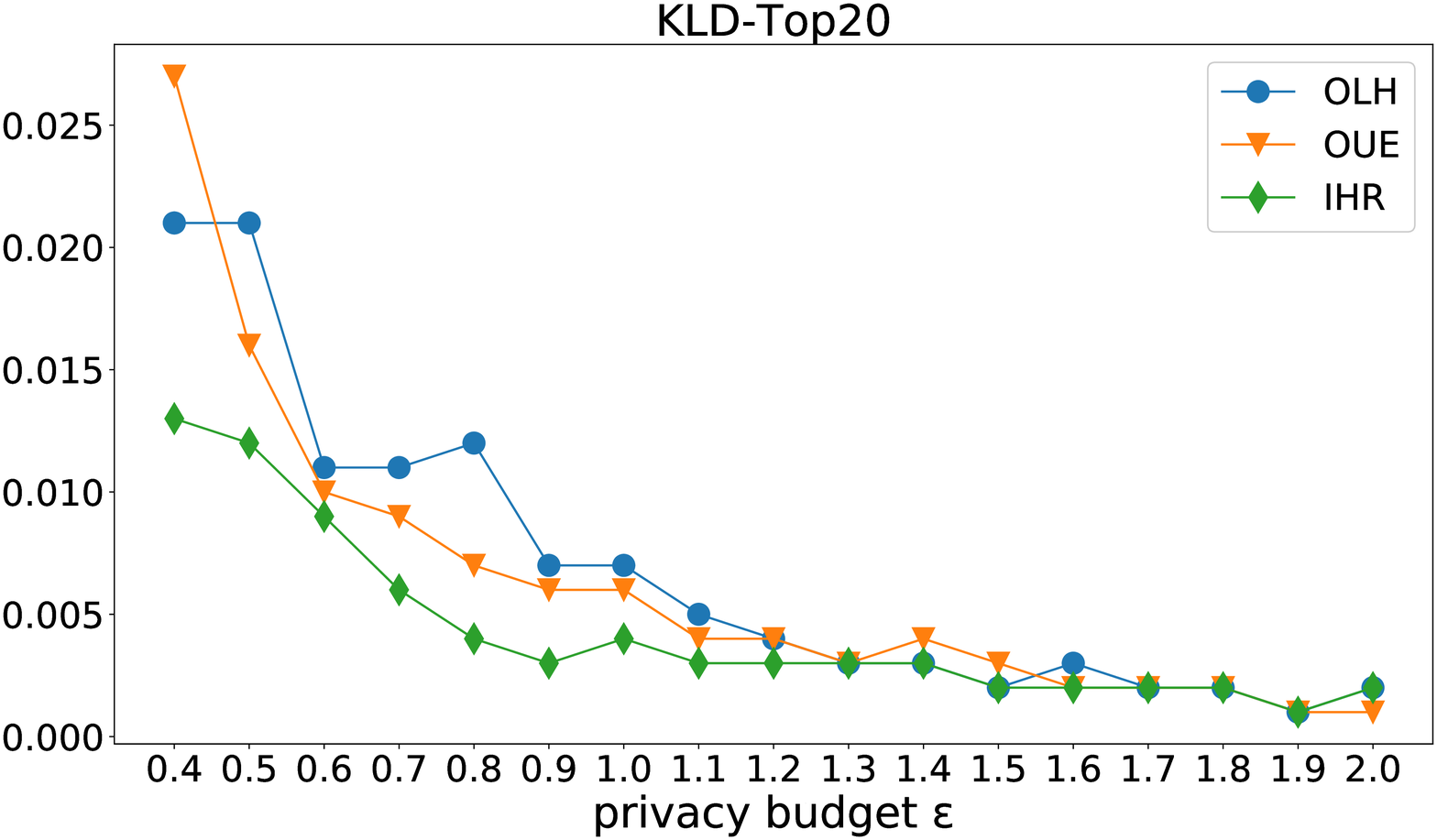}
		\label{fig:Online_Top20_KLD}}
	\subfigure[Cohort for Top 20]{
		\includegraphics[width=0.29\textwidth]{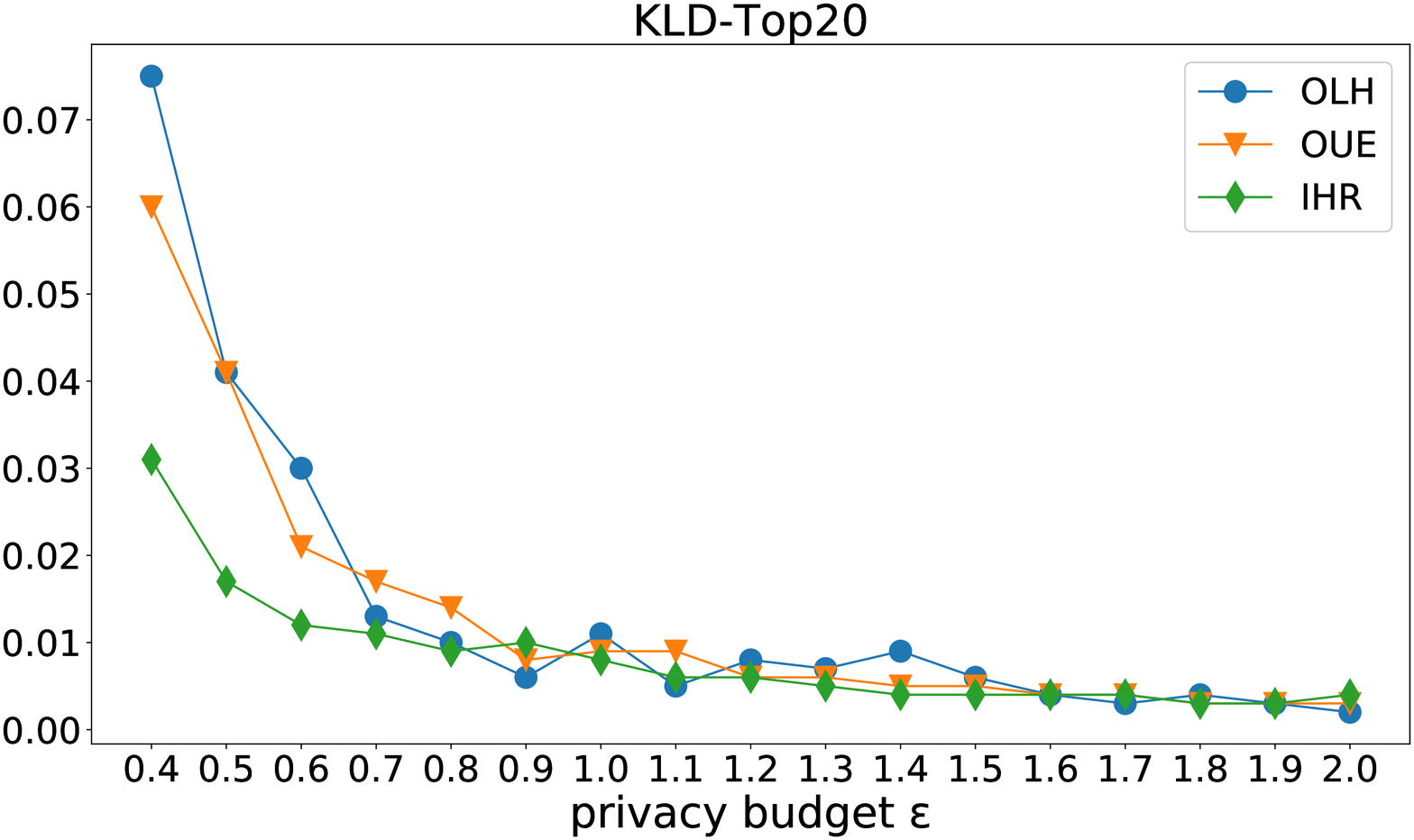}
		\label{fig:Cohort_Top20_KLD}}
	\subfigure[ZipfData for Top 50]{
		\includegraphics[width=0.29\textwidth]{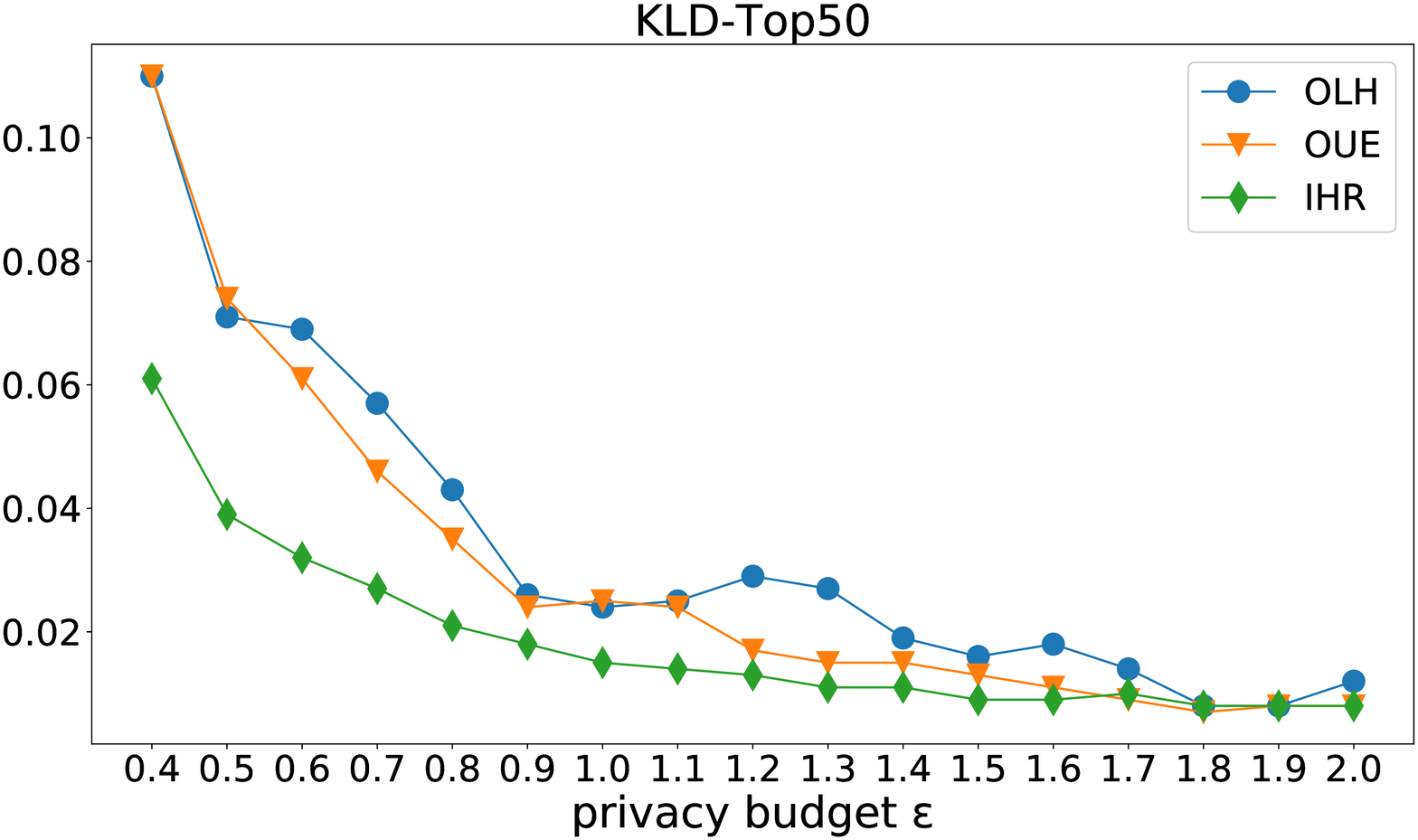}
		\label{fig:ZipfData_Top50_KLD}}
	\subfigure[Online for Top 50]{
		\includegraphics[width=0.29\textwidth]{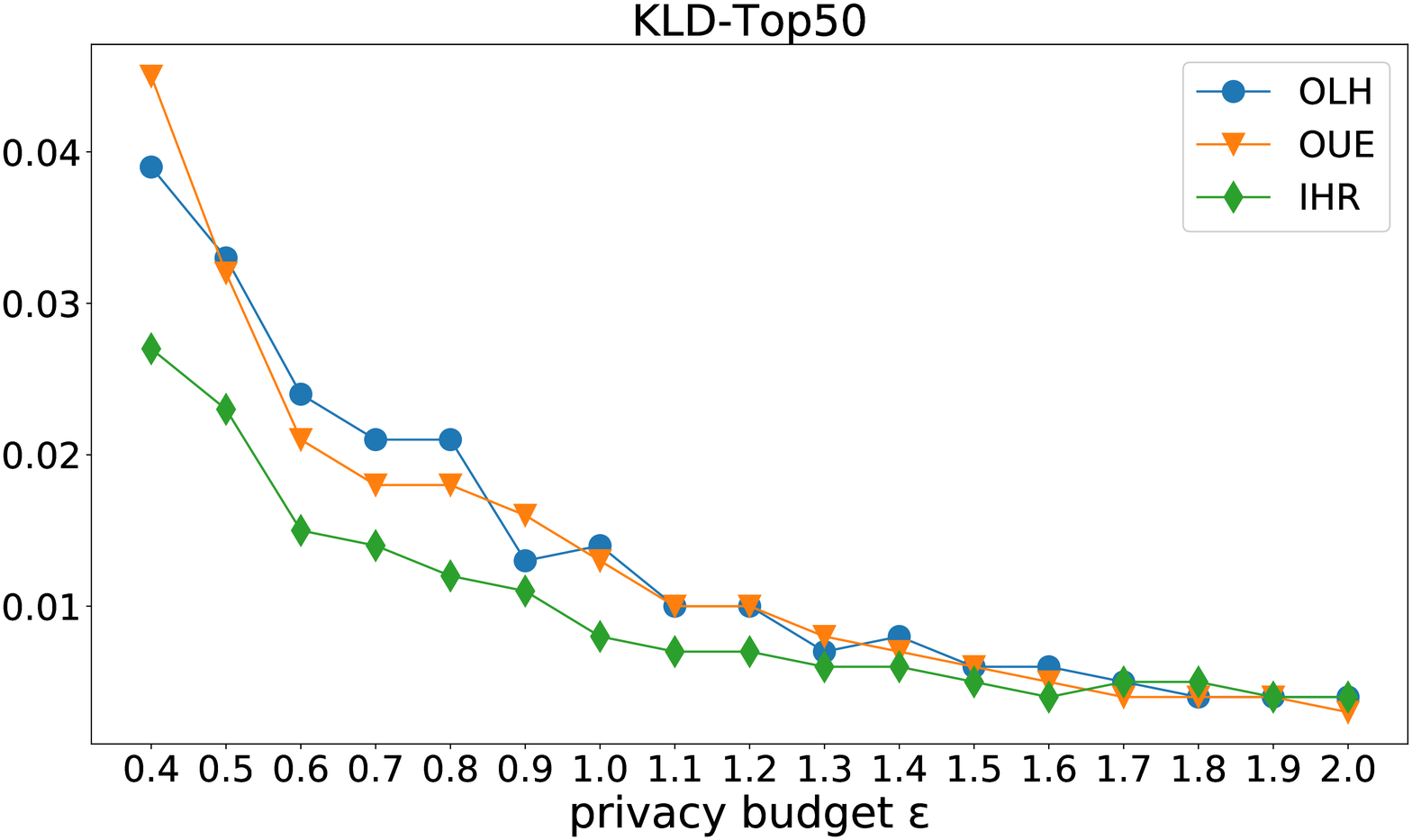}
		\label{fig:Online_Top50_KLD}}
	\subfigure[Cohort for Top 50]{
		\includegraphics[width=0.29\textwidth]{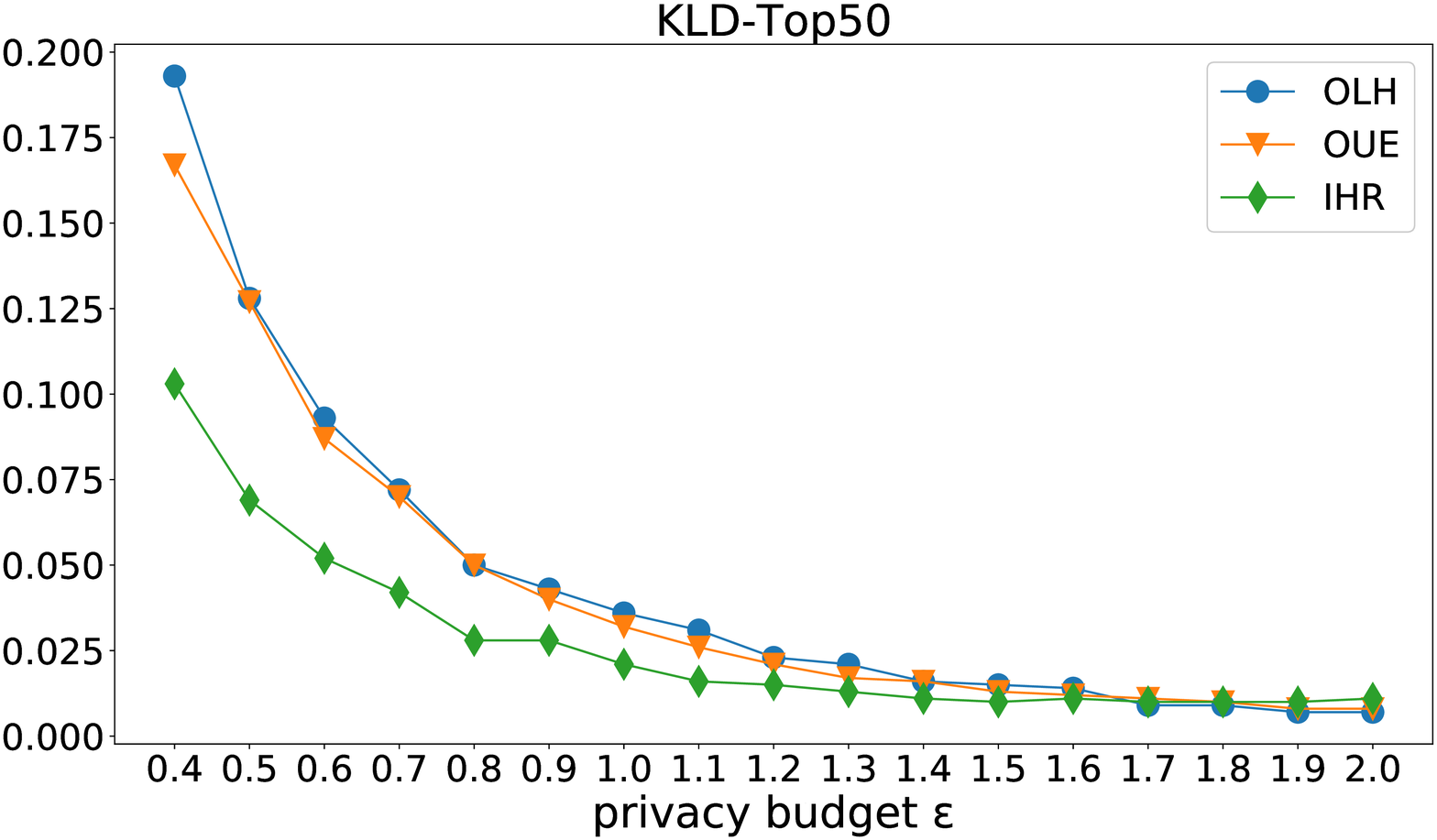}
		\label{fig:Cohort_Top50_KLD}}
	\subfigure[ZipfData for Top 100]{
		\includegraphics[width=0.29\textwidth]{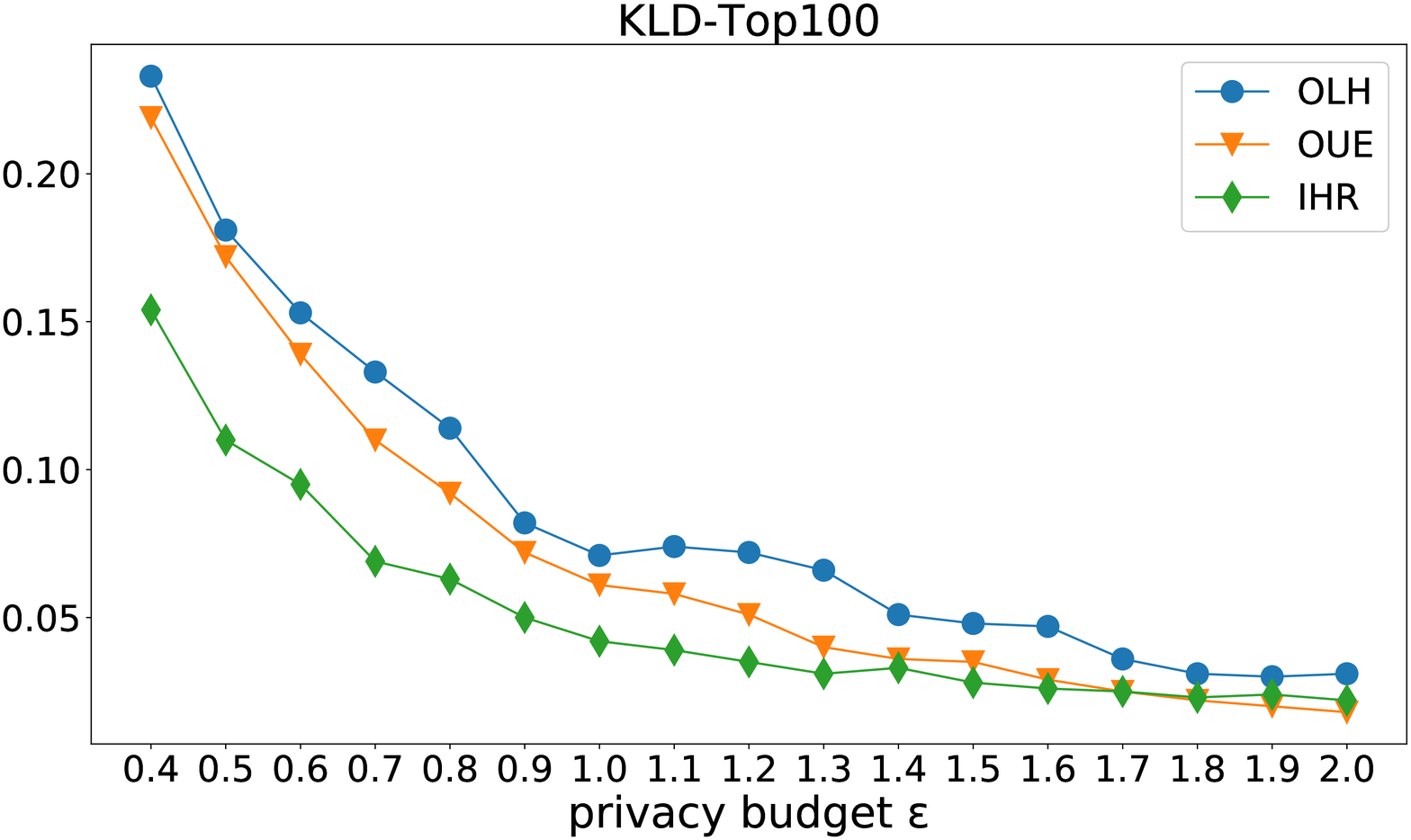}
		\label{fig:ZipfData_Top100_KLD}}
	\subfigure[Online for Top 100]{
		\includegraphics[width=0.29\textwidth]{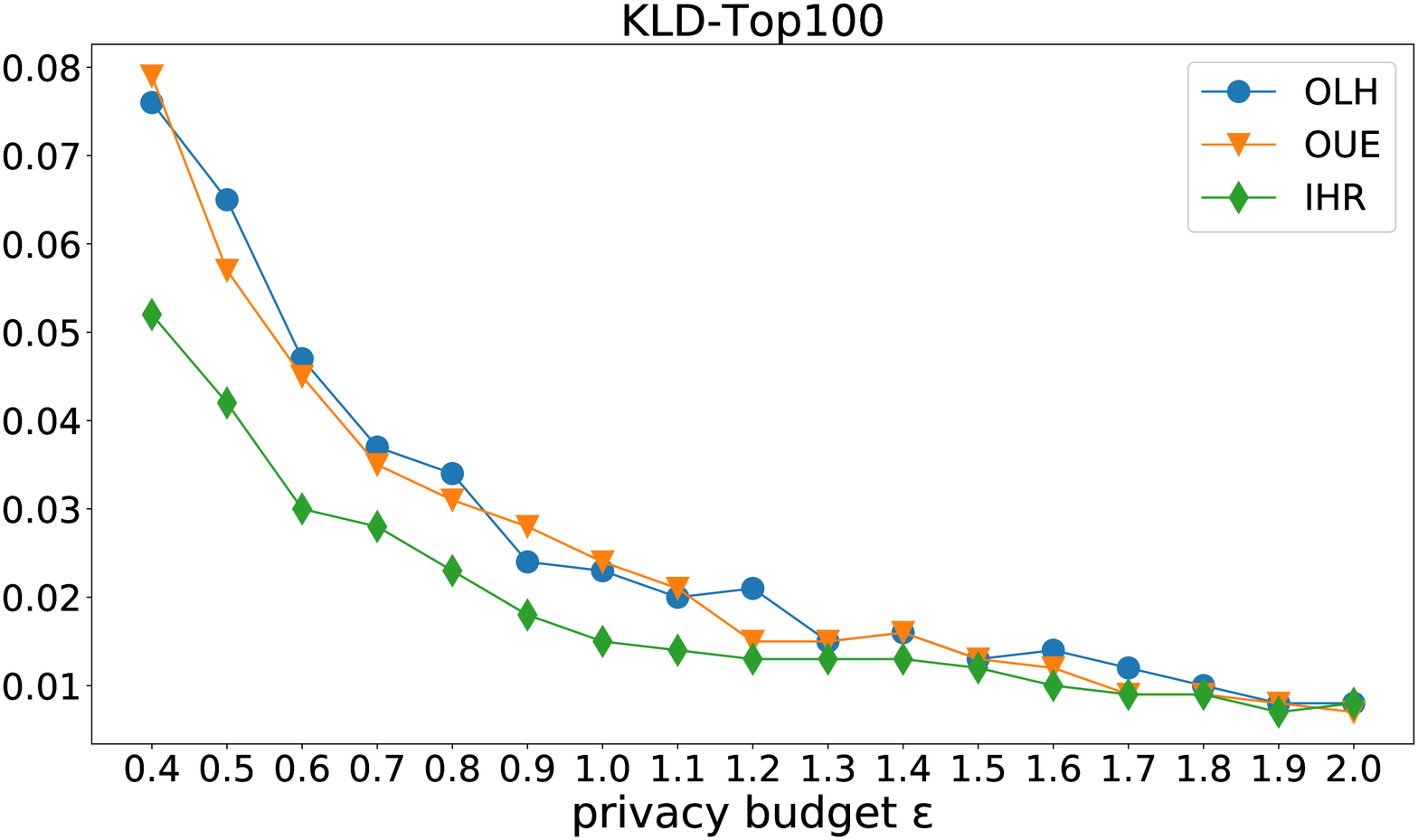}
		\label{fig:Online_Top100_KLD}}
	\subfigure[Cohort for Top 100]{
		\includegraphics[width=0.29\textwidth]{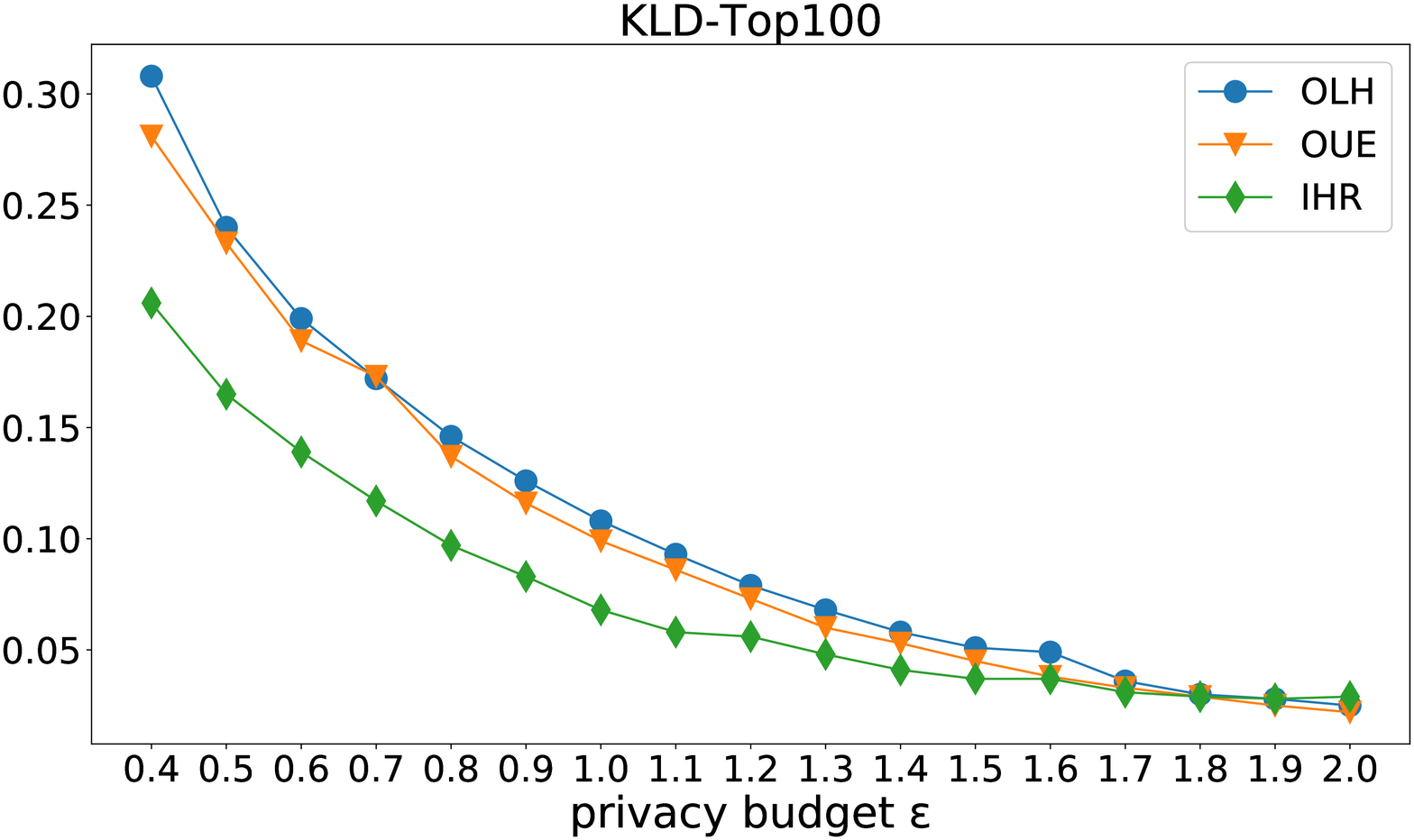}
		\label{fig:Cohort_Top100_KLD}}
	\caption{KLD for each dataset}
	\label{Fig: result_KLD}
\end{figure*}

First, Figure \ref{Fig: result_KLD} illustrates the KLD of FO on all datasets. In particular, the first row in Figures \ref{fig:ZipfData_Top20_KLD}, \ref{fig:Online_Top20_KLD}, and \ref{fig:Cohort_Top20_KLD}, the second row in Figures \ref{fig:ZipfData_Top50_KLD}, \ref{fig:Online_Top50_KLD}, and \ref{fig:Cohort_Top50_KLD}, and the third row in Figures \ref{fig:ZipfData_Top100_KLD}, \ref{fig:Online_Top100_KLD}, and \ref{fig:Cohort_Top100_KLD} present trendlines with increasing privacy budgets when the top 20, top 50, and top 100 values are selected, respectively.  The trendlines in these figures decrease as the privacy budget $\varepsilon$ increases, and the trendlines for FHR (green) are the lowest. According to the definition of KLD, the lower the trendline, the more accurate the FO. Thus, FHO performs best with respect to KLD.

\begin{figure*}[!htp]
	\centering
	\graphicspath{{Img/}}
	\subfigure[ZipfData for Top 20]{
		\includegraphics[width=0.29\textwidth]{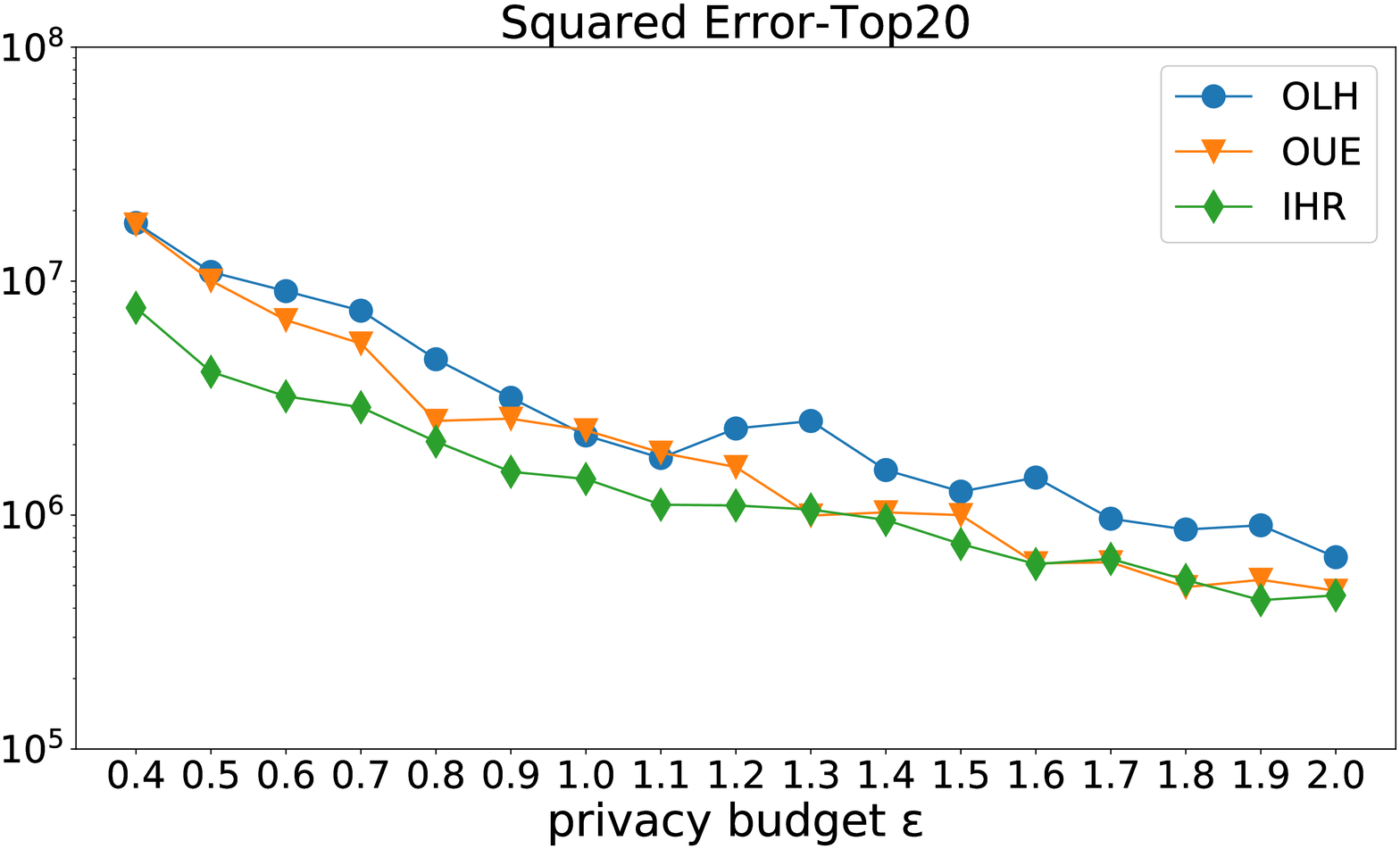}
		\label{fig:ZipfData_Top20_SE}}
	\subfigure[Online for Top 20]{
		\includegraphics[width=0.29\textwidth]{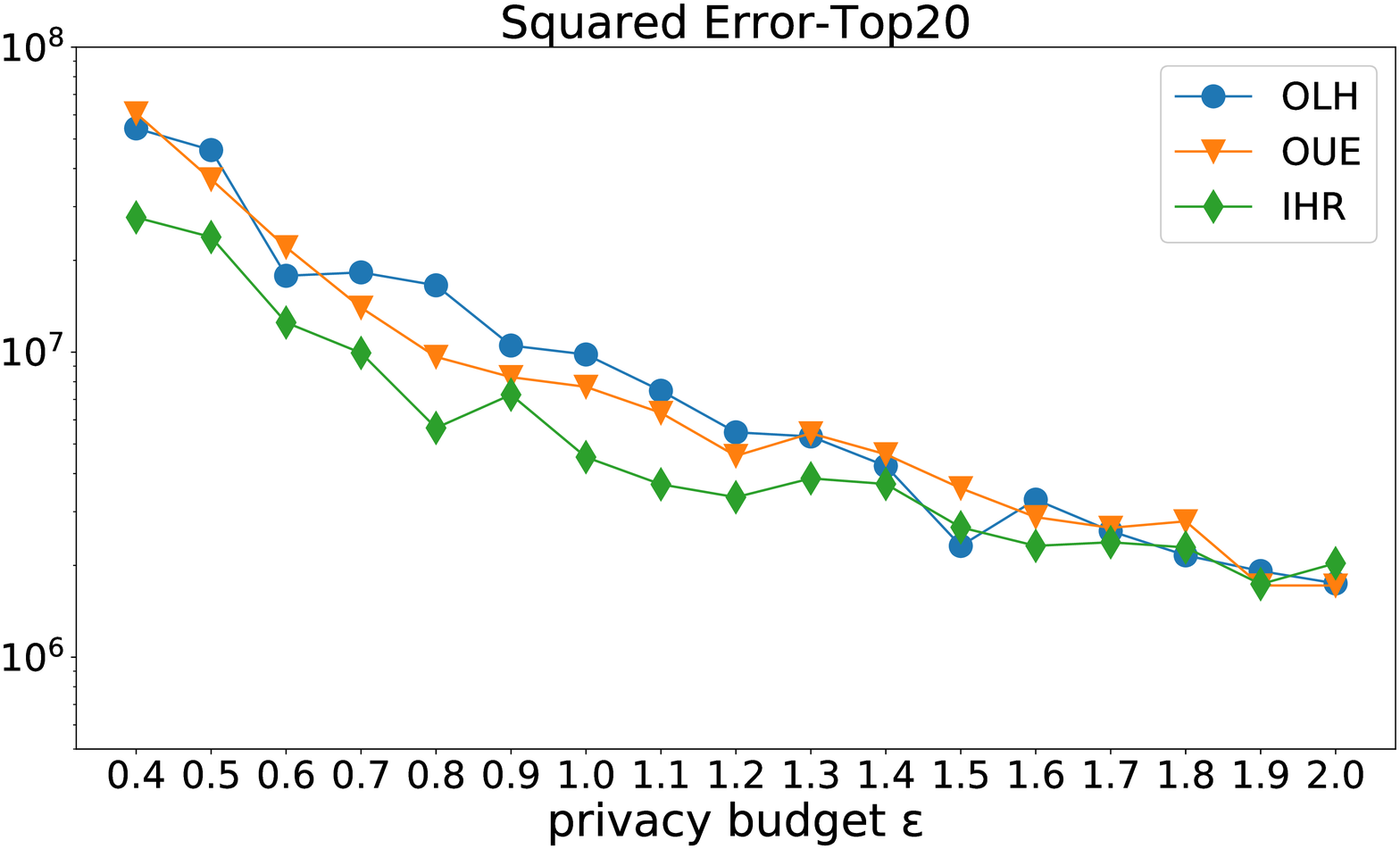}
		\label{fig:Online_Top20_SE}}
	\subfigure[Cohort for Top 20]{
		\includegraphics[width=0.29\textwidth]{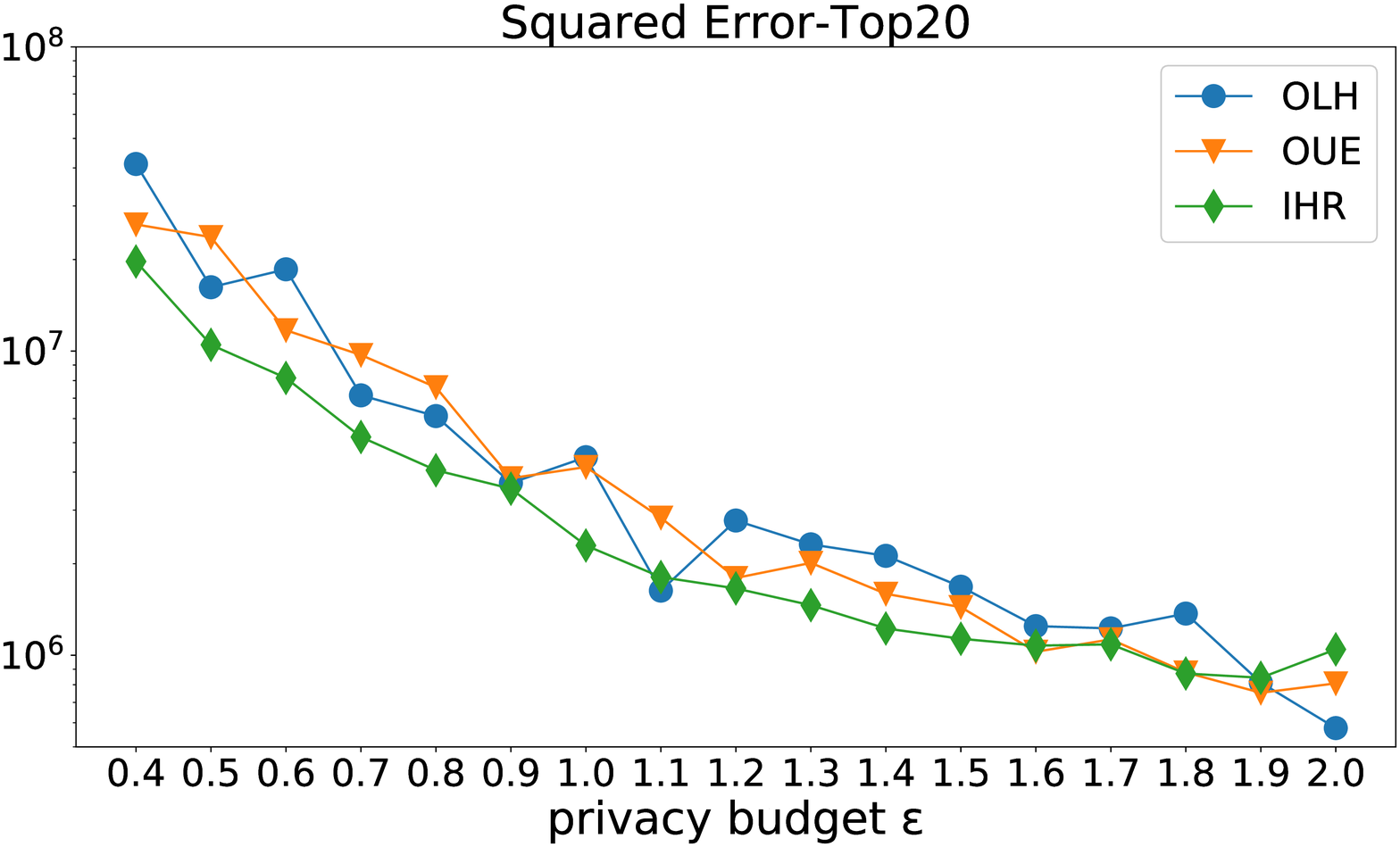}
		\label{fig:Cohort_Top20_SE}}
	\subfigure[ZipfData for Top 50]{
		\includegraphics[width=0.29\textwidth]{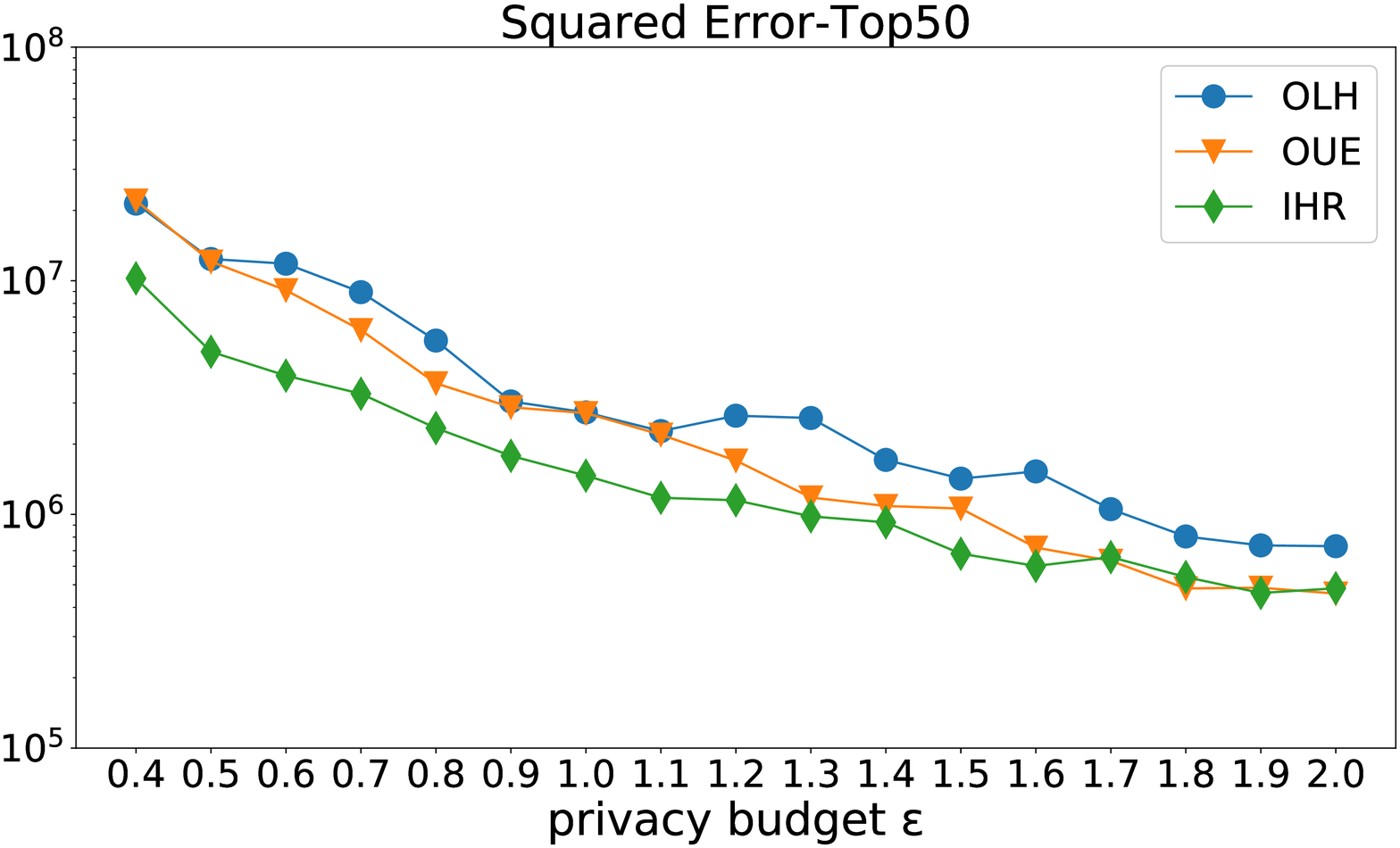}
		\label{fig:ZipfData_Top50_SE}}
	\subfigure[Online for Top 50]{
		\includegraphics[width=0.29\textwidth]{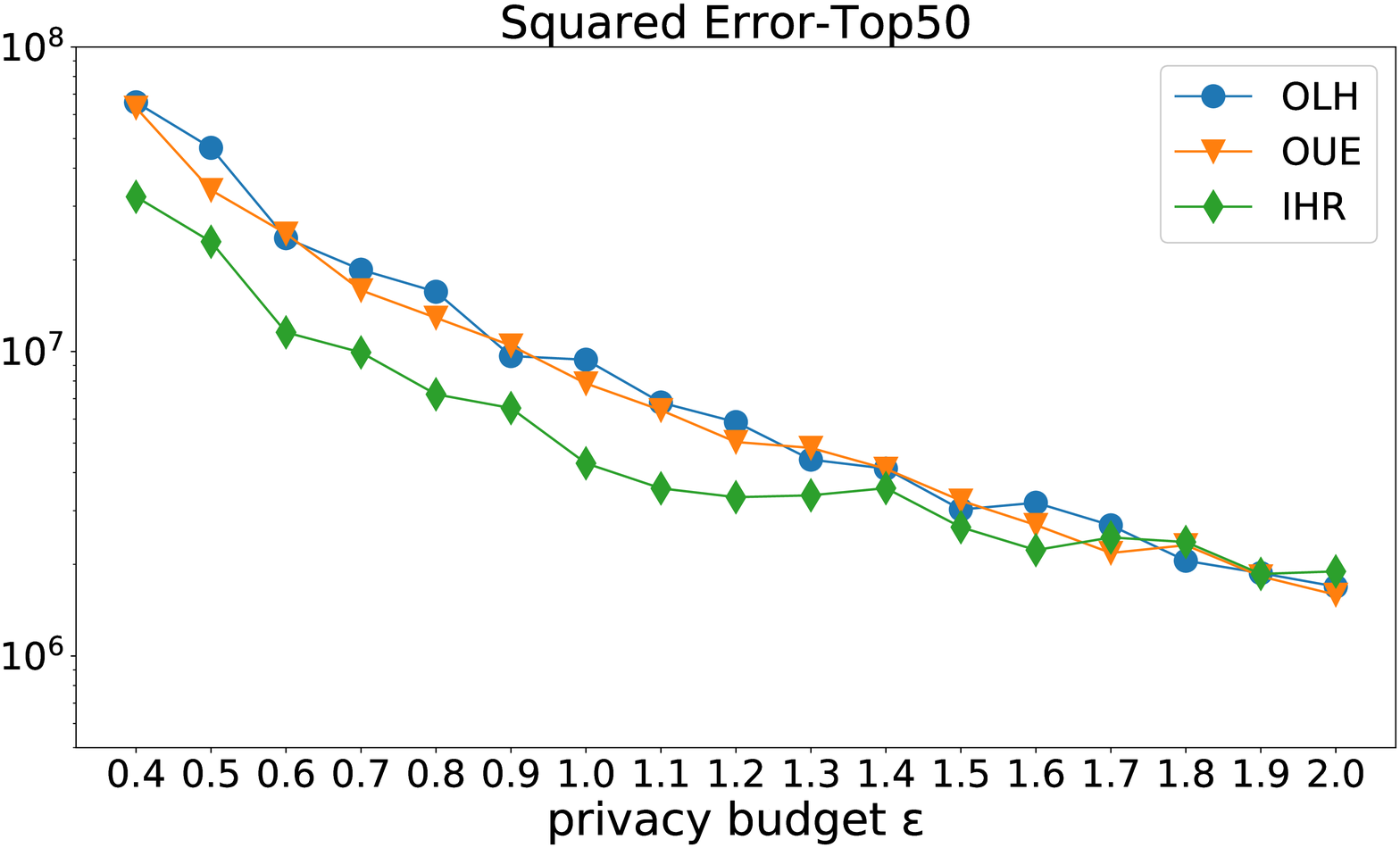}
		\label{fig:Online_Top50_SE}}
	\subfigure[Cohort for Top 50]{
		\includegraphics[width=0.29\textwidth]{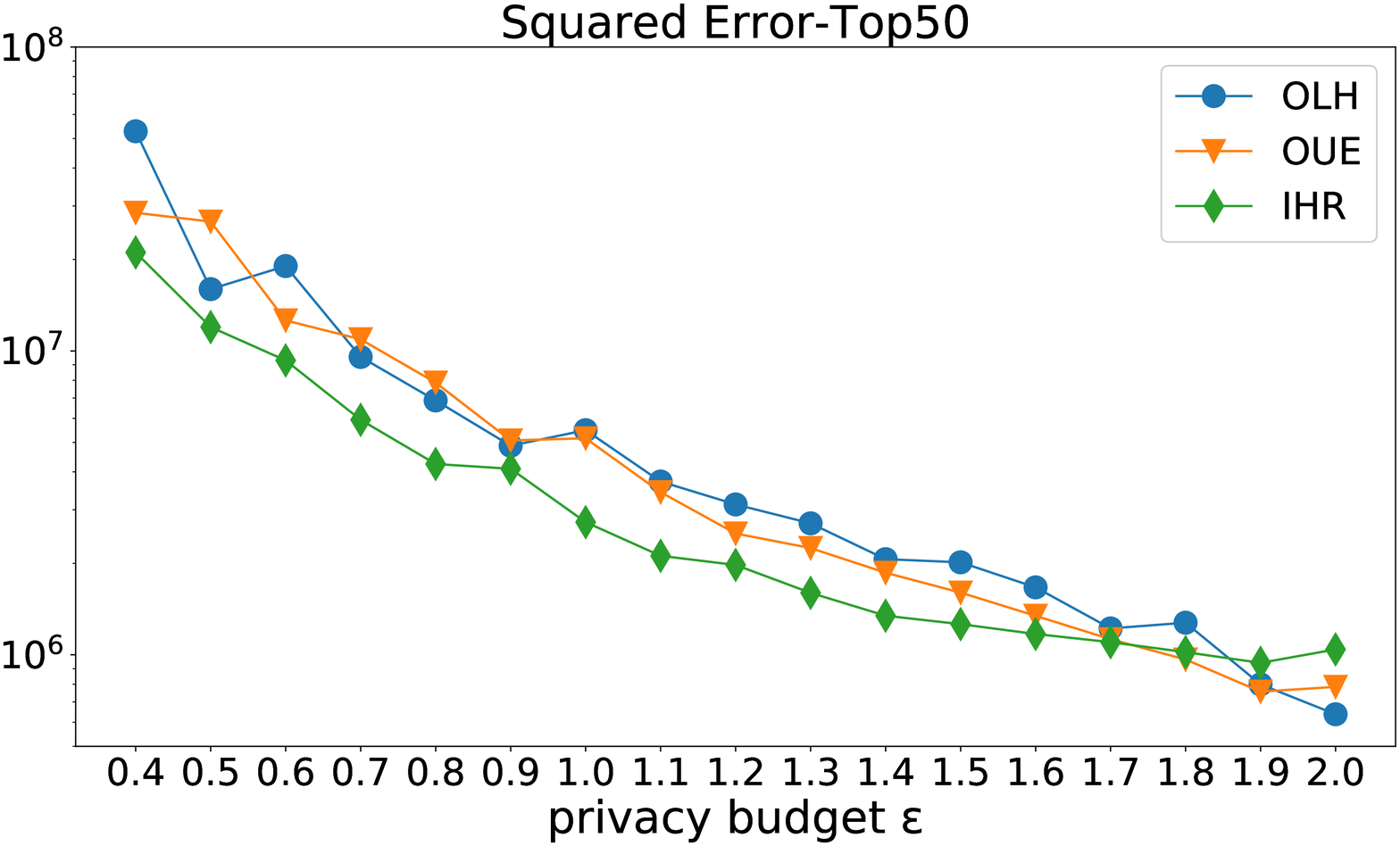}
		\label{fig:Cohort_Top50_SE}}
	\subfigure[ZipfData for Top 100]{
		\includegraphics[width=0.29\textwidth]{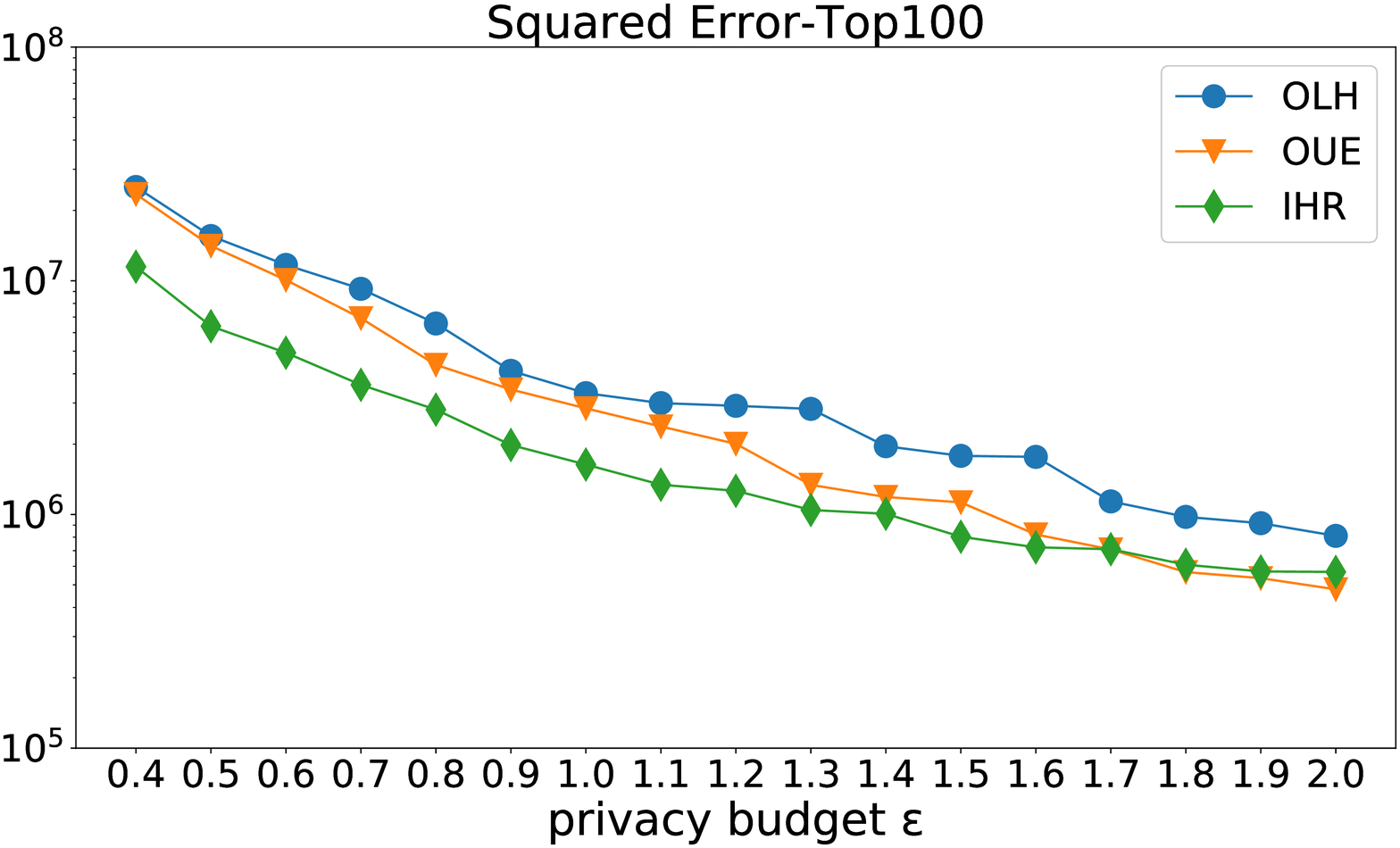}
		\label{fig:ZipfData_Top100_SE}}
	\subfigure[Online for Top 100]{
		\includegraphics[width=0.29\textwidth]{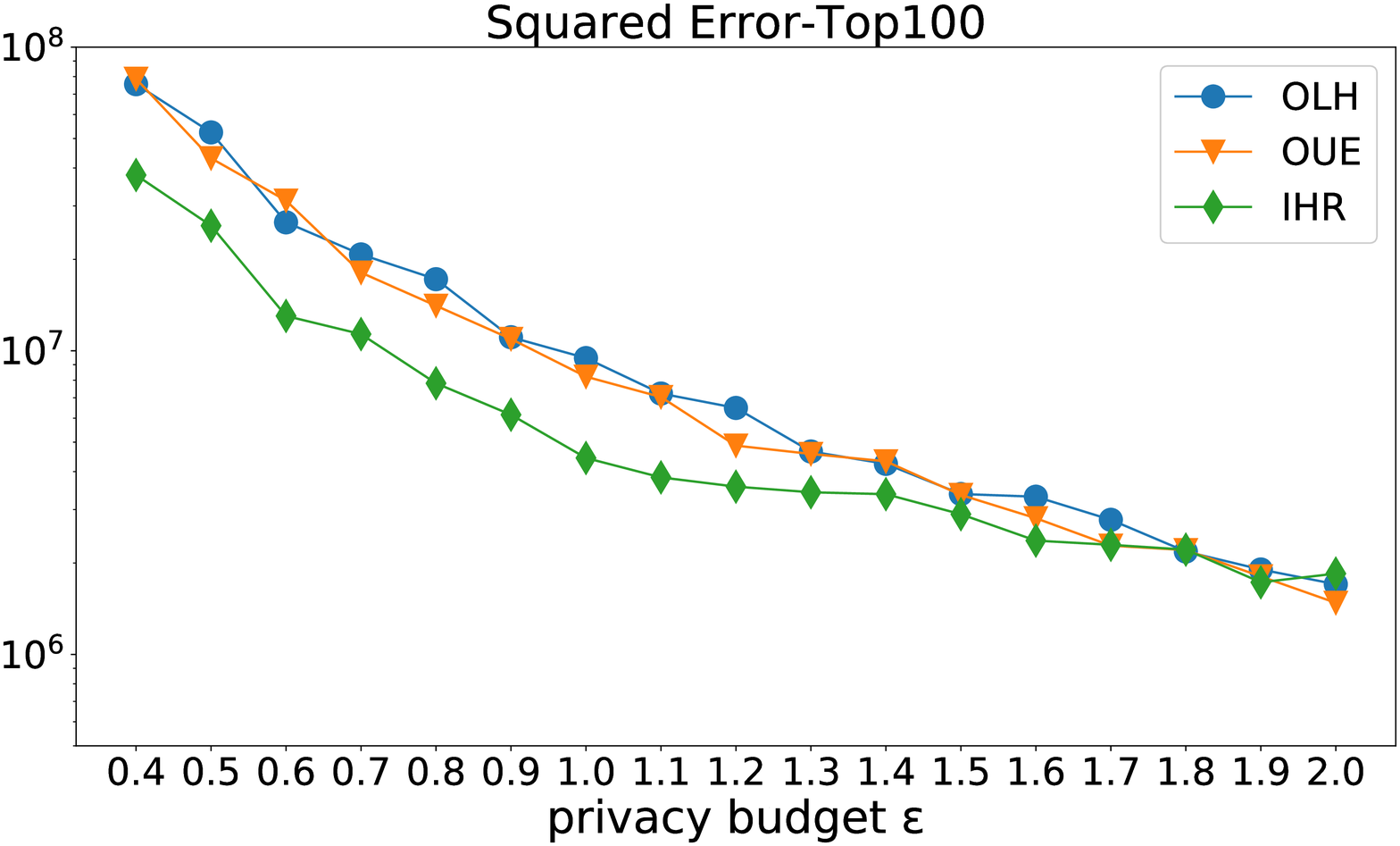}
		\label{fig:Online_Top100_SE}}
	\subfigure[Cohort for Top 100]{
		\includegraphics[width=0.29\textwidth]{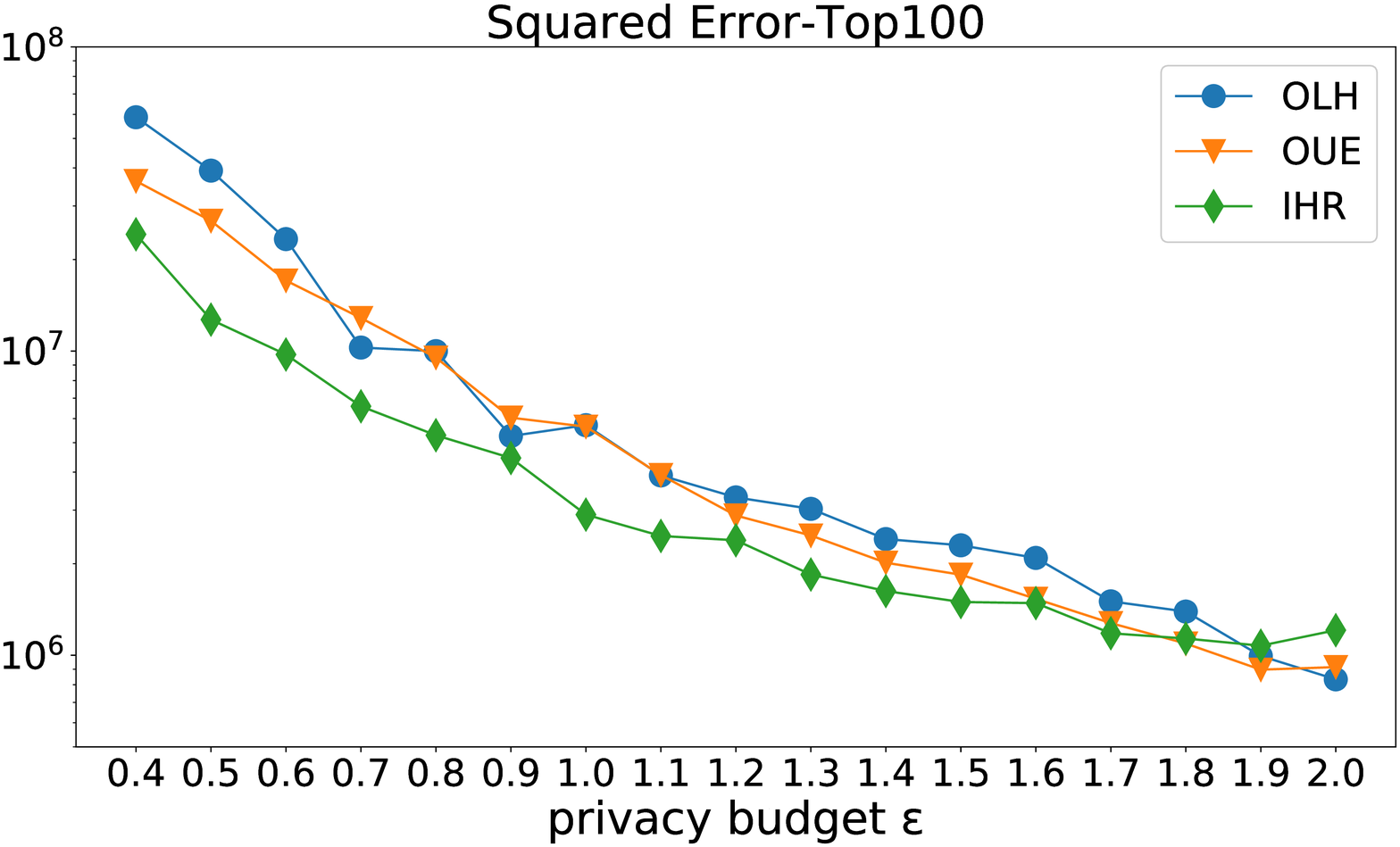}
		\label{fig:Cohort_Top100_SE}}
	
	\caption{Squared error for each dataset}
	\label{Fig: result_SE}
\end{figure*}

Secongd, Figure \ref{Fig: result_SE} presents the squared error (SE) for the FO for all datasets. In particular, the first row in Figures \ref{fig:ZipfData_Top20_SE} \ref{fig:Online_Top20_SE}, and \ref{fig:Cohort_Top20_SE}, second row in Figures \ref{fig:ZipfData_Top50_SE}, \ref{fig:Online_Top50_SE}, and \ref{fig:Cohort_Top50_SE}, and third row in Figures \ref{fig:ZipfData_Top100_SE}, \ref{fig:Online_Top100_SE}, and \ref{fig:Cohort_Top100_SE} show trendlines with increasing privacy budgets when the top 20, top 50 and top 100 values are selected, respectively.  The trendlines decrease as the privacy budget $\varepsilon$ increases, and the trendlines for FHR (green) are the lowest overall. According to the definition of SE, the lower the trendline, the more accurate the FO’s frequencies . Thus, the mechanism FHO performs best with respect to SE.

\begin{figure*}[!htp]
	\centering
	\graphicspath{{Img/}}
	\subfigure[ZipfData for Top 20]{
		\includegraphics[width=0.29\textwidth]{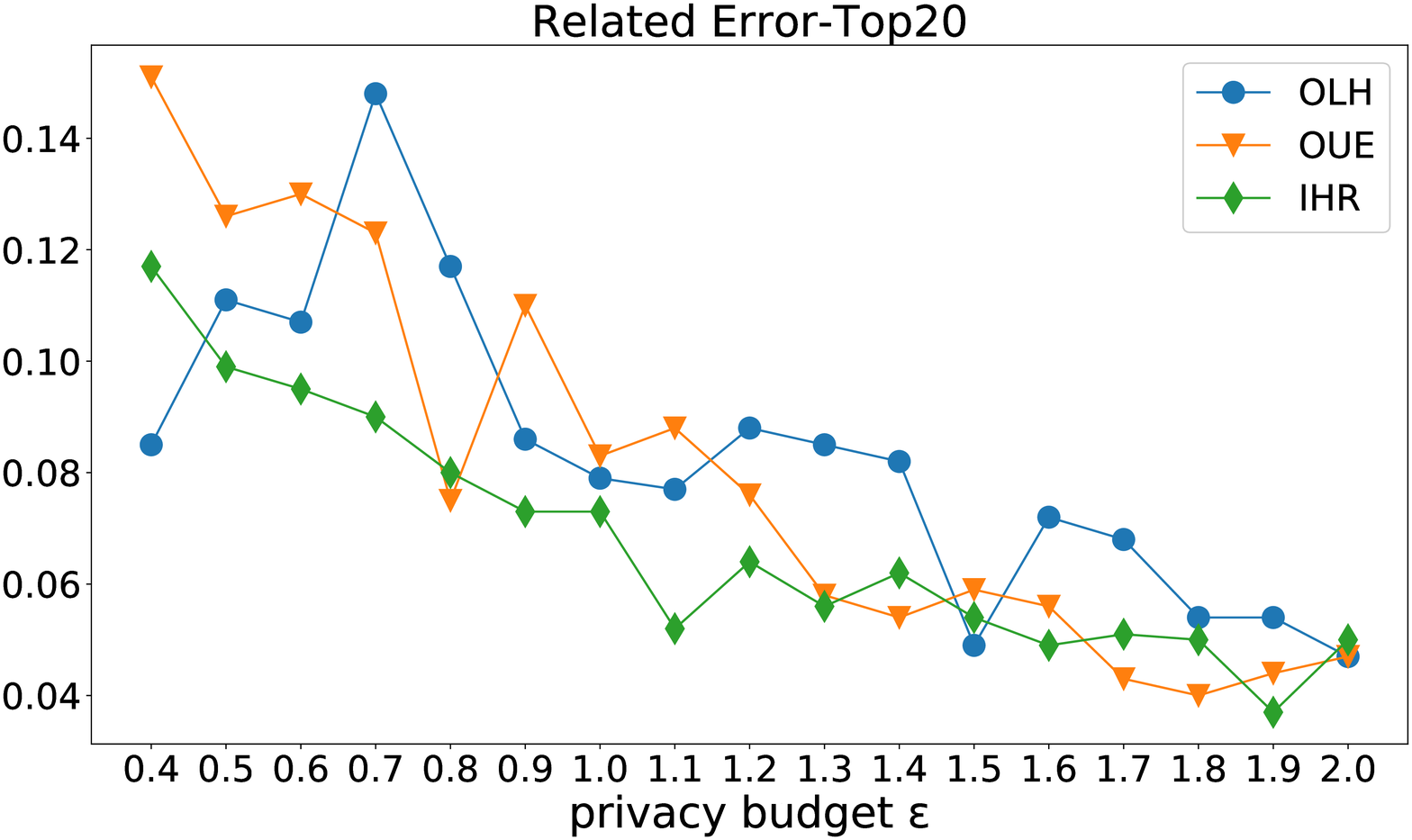}
		\label{fig:ZipfData_Top20_RE}}
	\subfigure[Online for Top 20]{
		\includegraphics[width=0.29\textwidth]{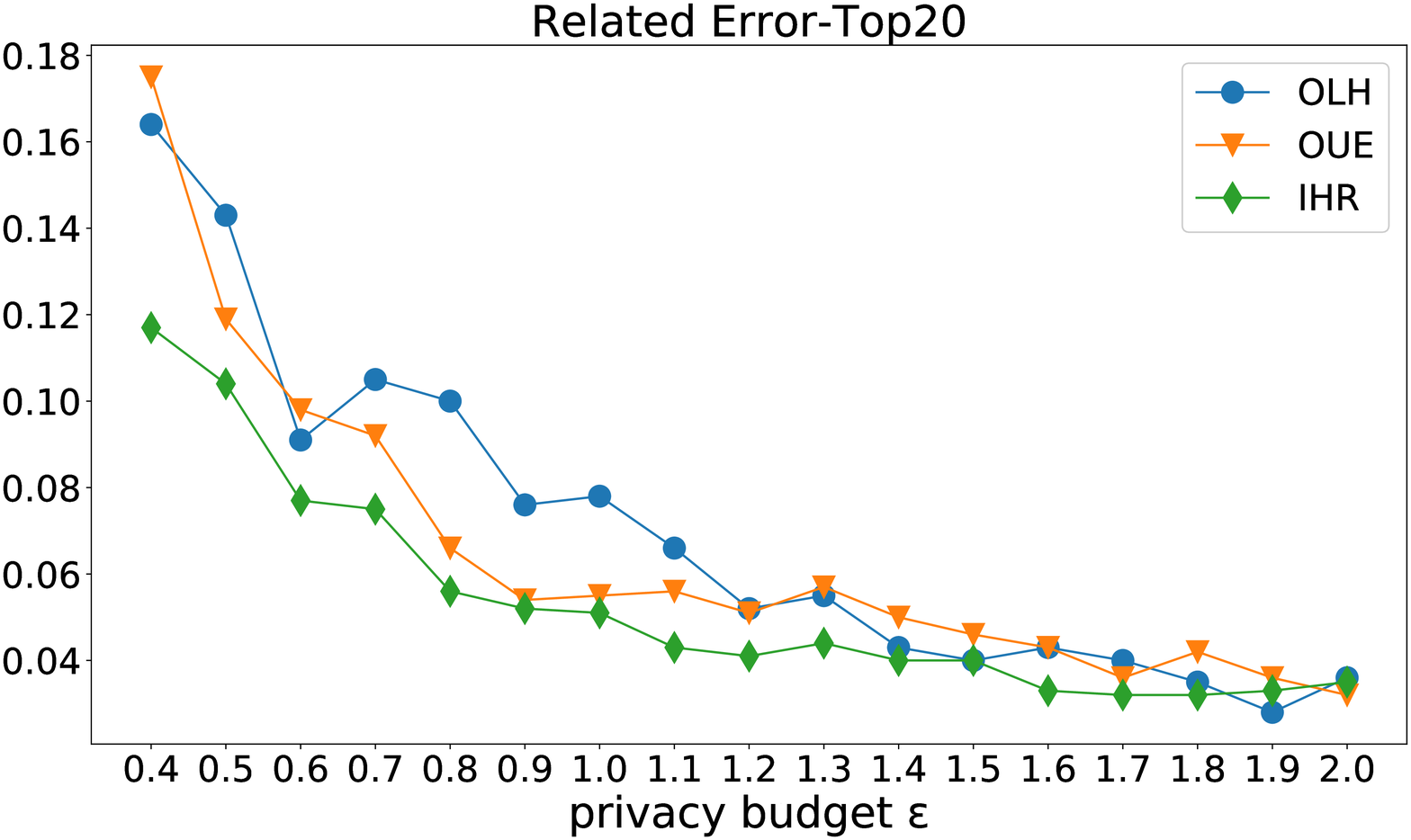}
		\label{fig:Online_Top20_RE}}
	\subfigure[Cohort for Top 20]{
		\includegraphics[width=0.29\textwidth]{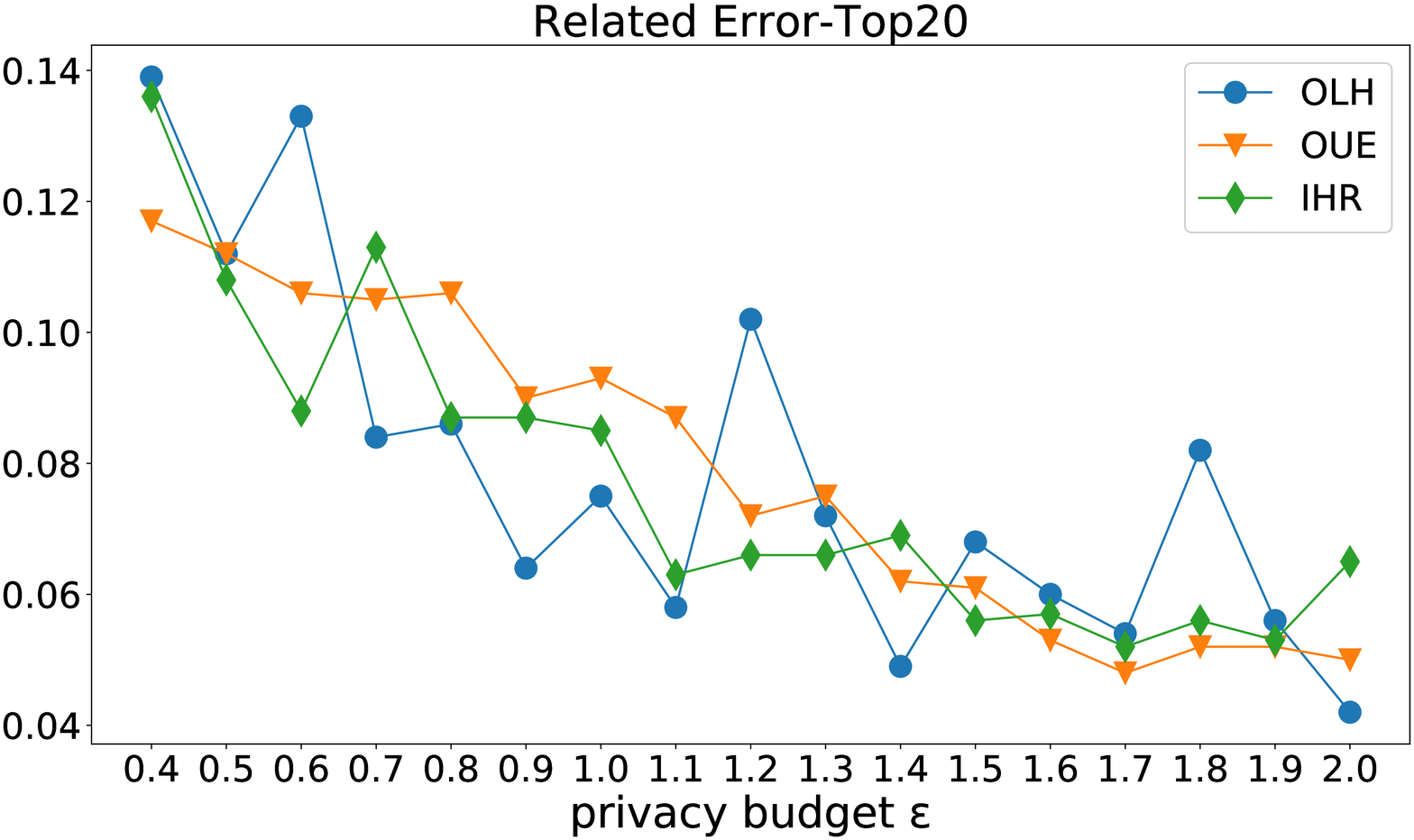}
		\label{fig:Cohort_Top20_RE}}
	\subfigure[ZipfData for Top 50]{
		\includegraphics[width=0.29\textwidth]{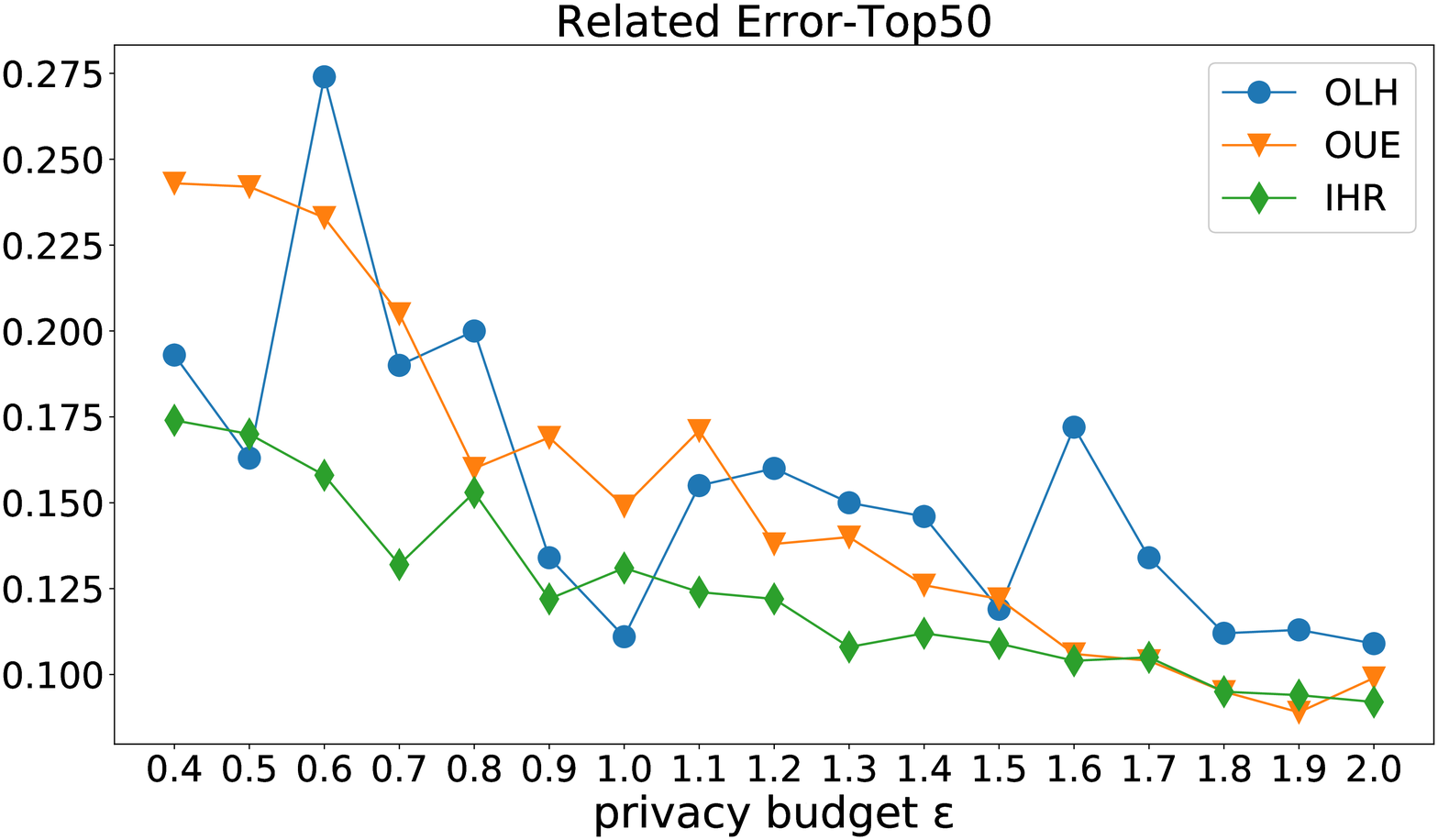}
		\label{fig:ZipfData_Top50_RE}}
	\subfigure[Online for Top 50]{
		\includegraphics[width=0.29\textwidth]{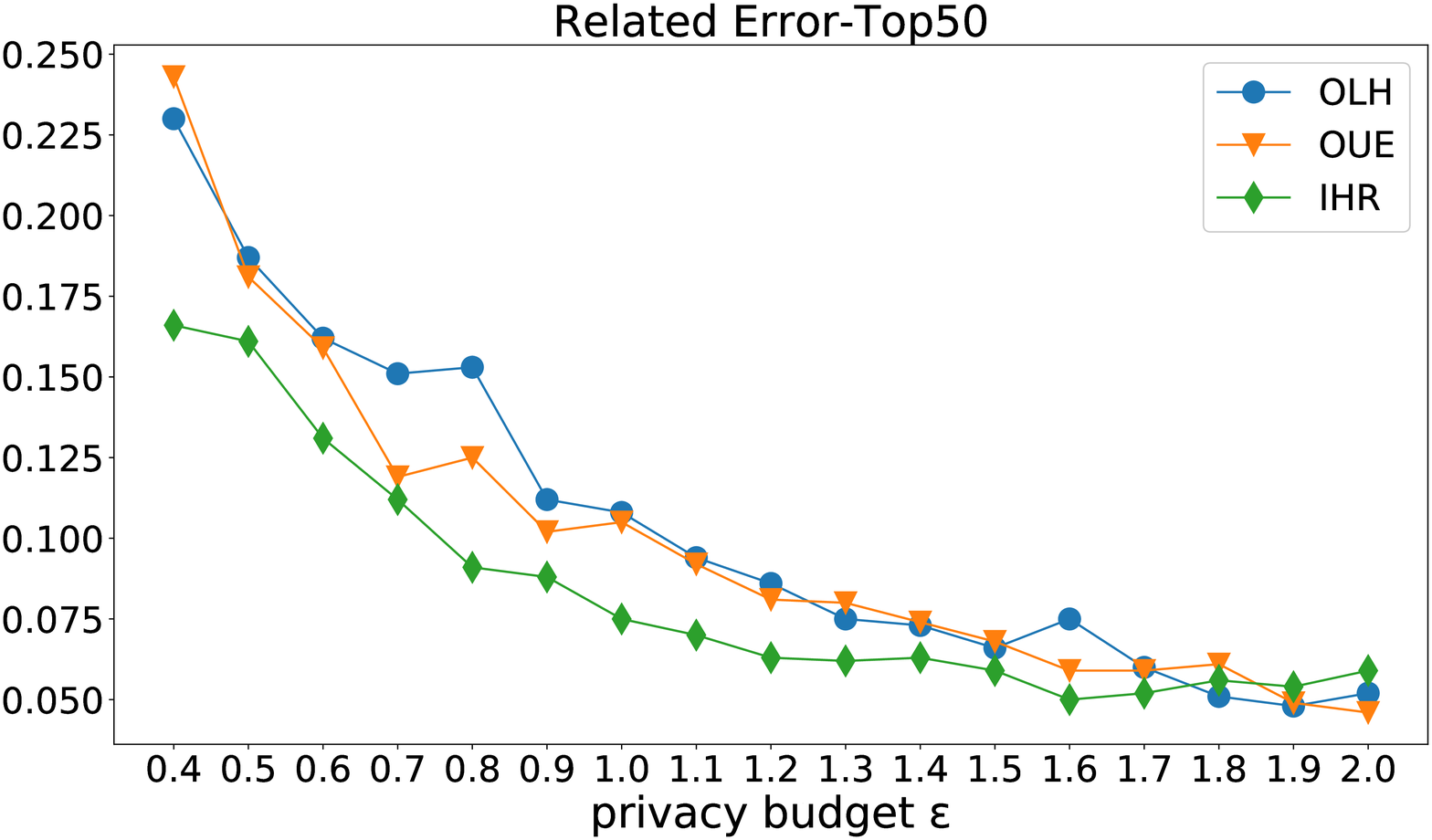}
		\label{fig:Online_Top50_RE}}
	\subfigure[Cohort for Top 50]{
		\includegraphics[width=0.29\textwidth]{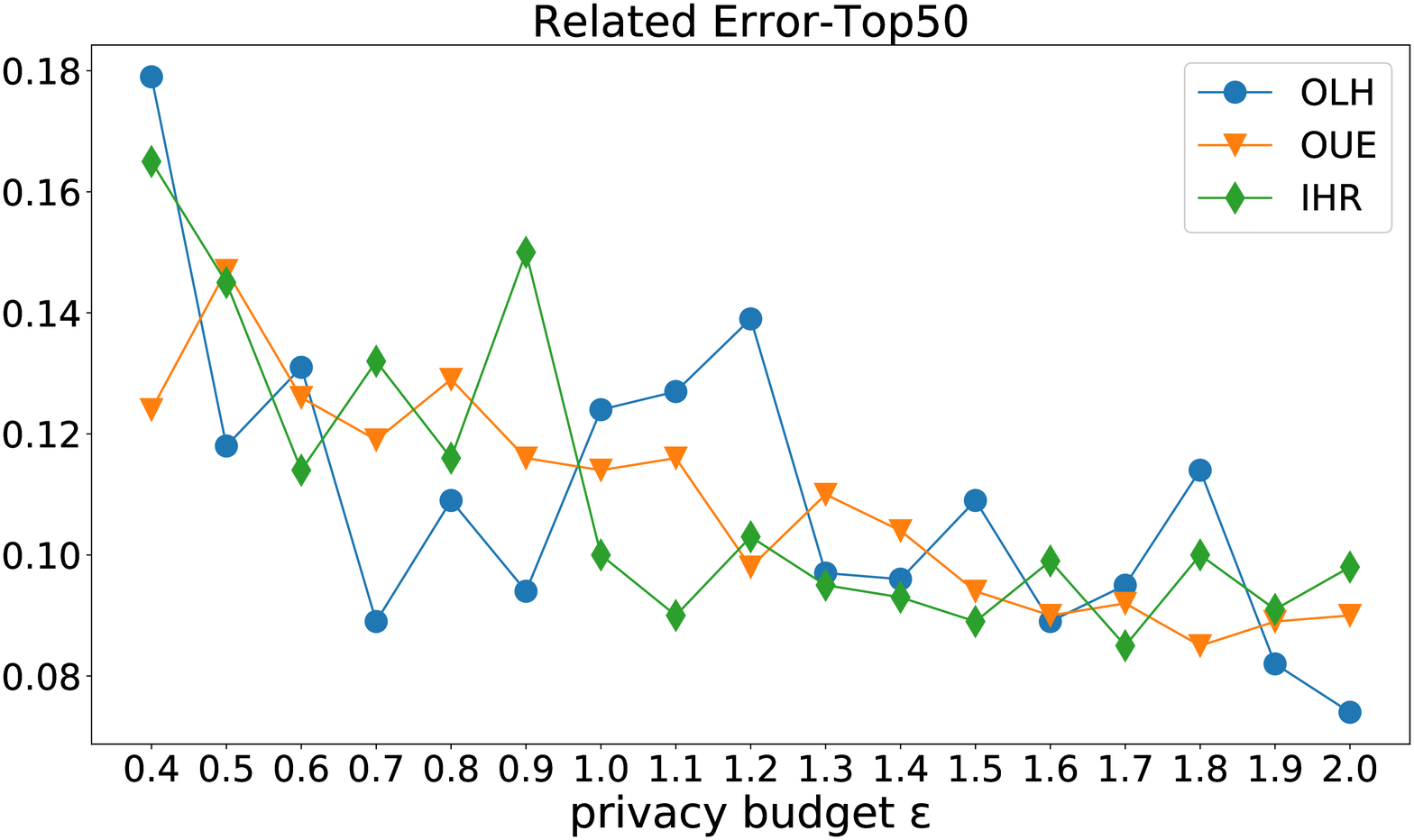}
		\label{fig:Cohort_Top50_RE}}
	\subfigure[ZipfData for Top 100]{
		\includegraphics[width=0.29\textwidth]{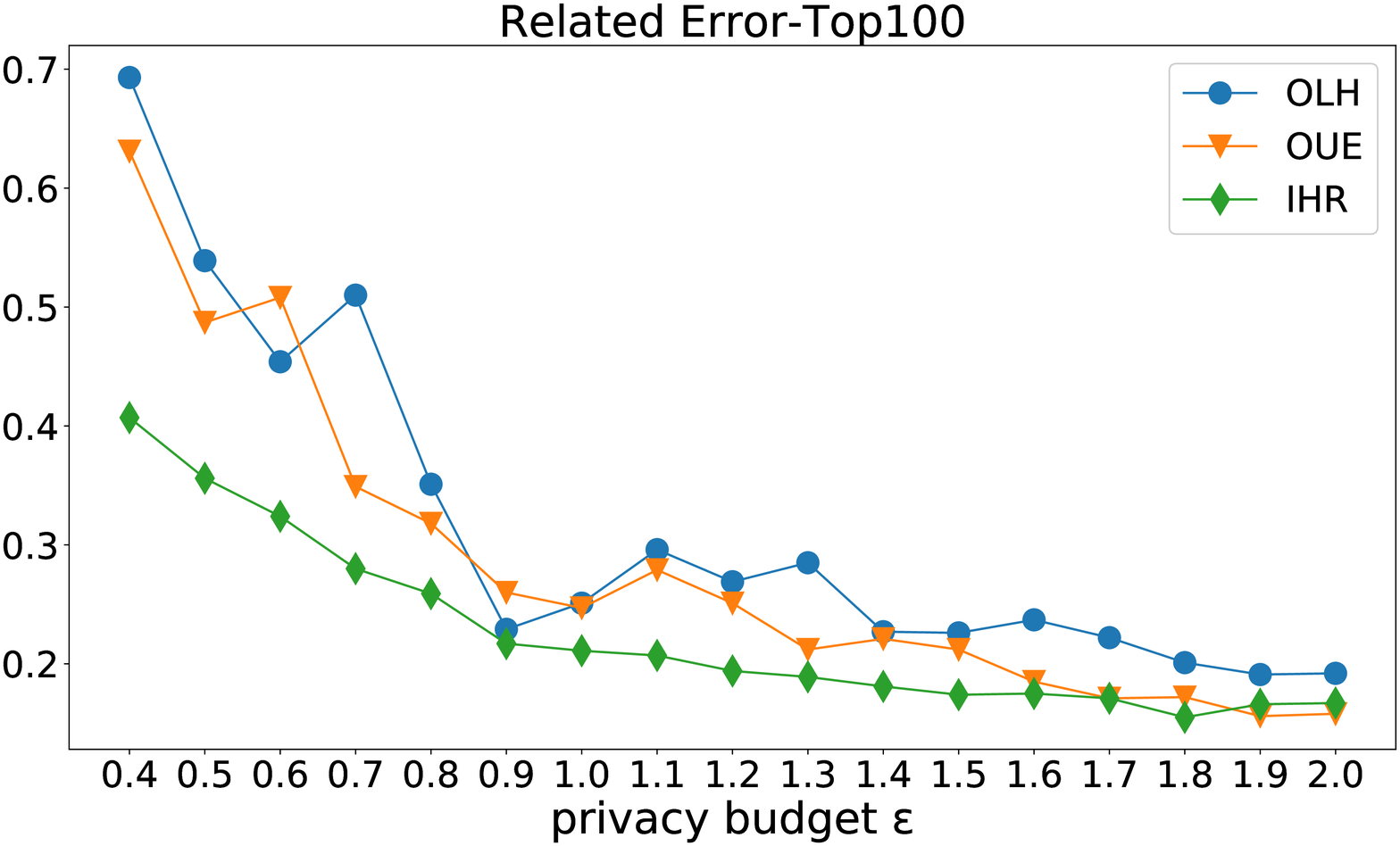}
		\label{fig:ZipfData_Top100_RE}}
	\subfigure[Online for Top 100]{
		\includegraphics[width=0.29\textwidth]{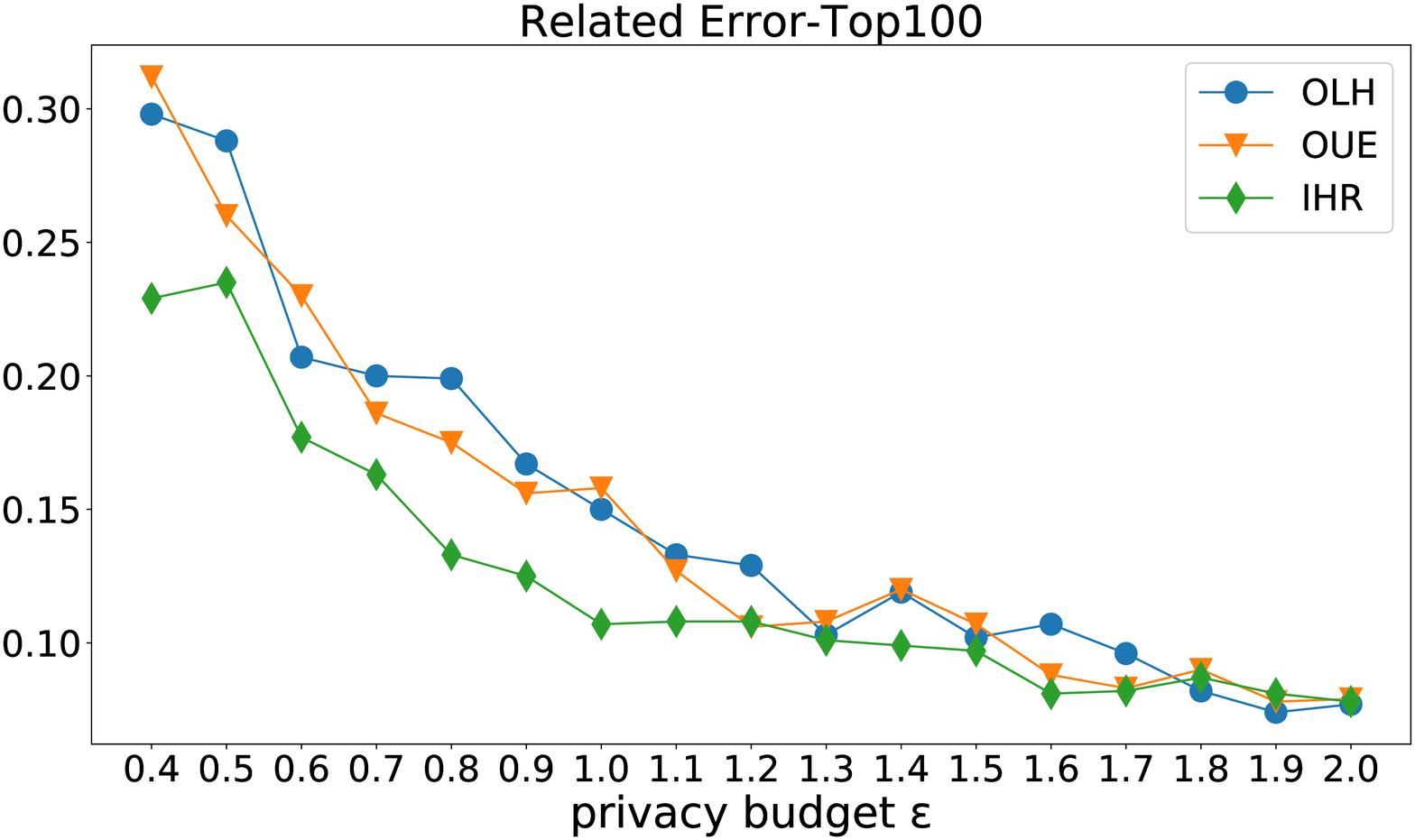}
		\label{fig:Online_Top100_RE}}
	\subfigure[Cohort for Top 100]{
		\includegraphics[width=0.29\textwidth]{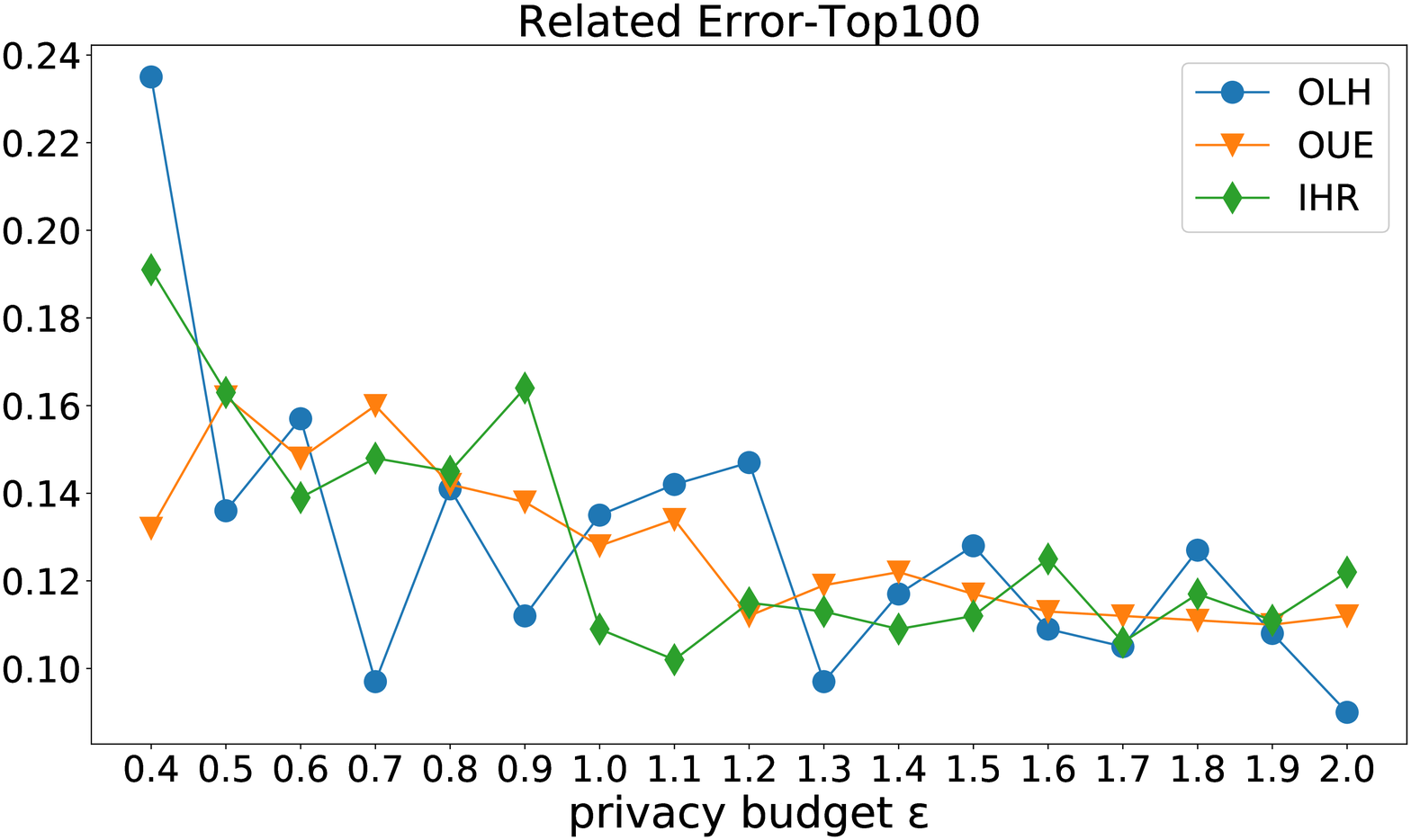}
		\label{fig:Cohort_Top100_RE}}
	\caption{Related Error for each dataset}
	\label{Fig: result_RE}
\end{figure*}

\begin{figure*}[htp]
	\centering
	\graphicspath{{Img/}}
	\subfigure[ZipfData for Top 20]{
		\includegraphics[width=0.29\textwidth]{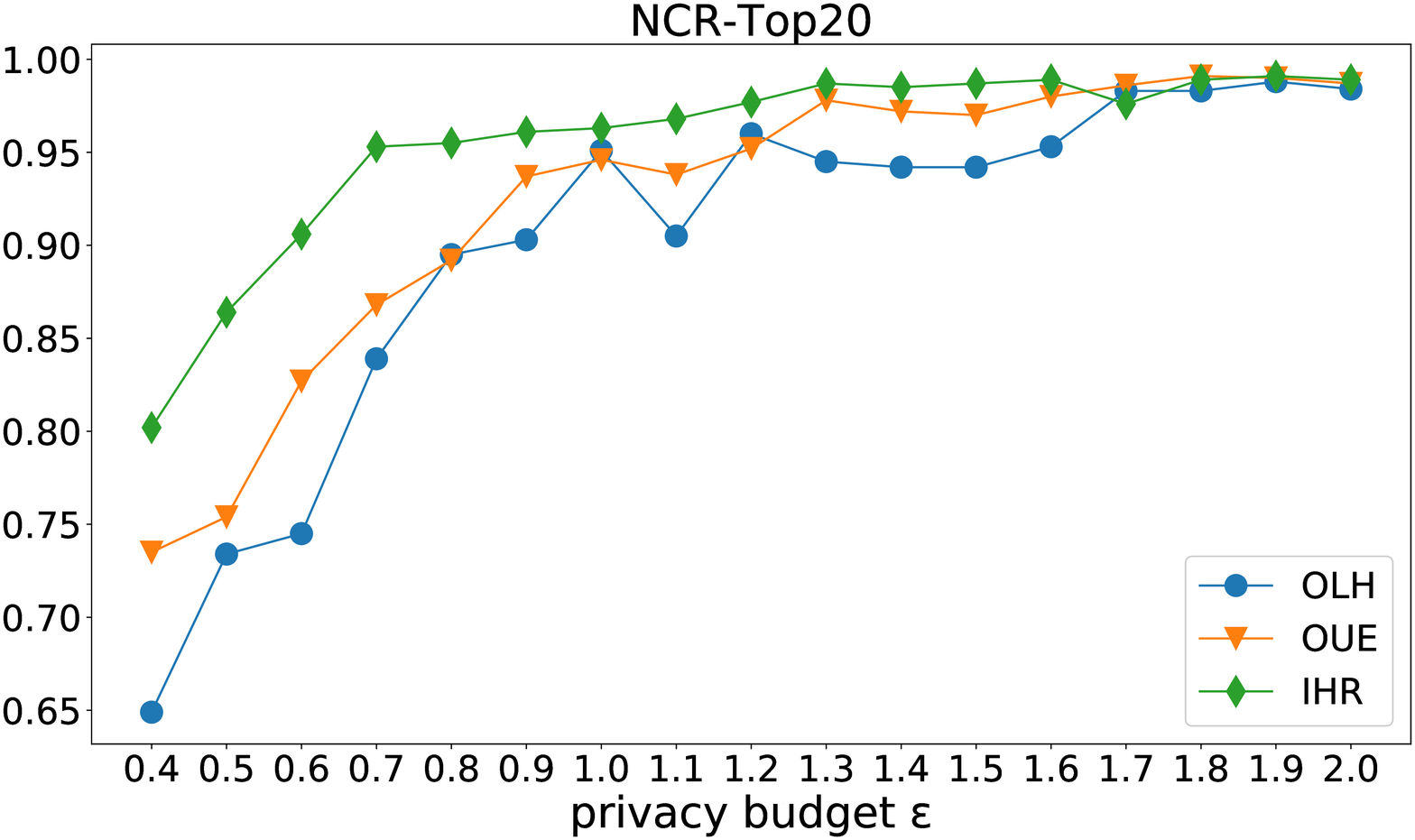}
		\label{fig:ZipfData_Top20_NCR}}
	\subfigure[Online for Top 20]{
		\includegraphics[width=0.29\textwidth]{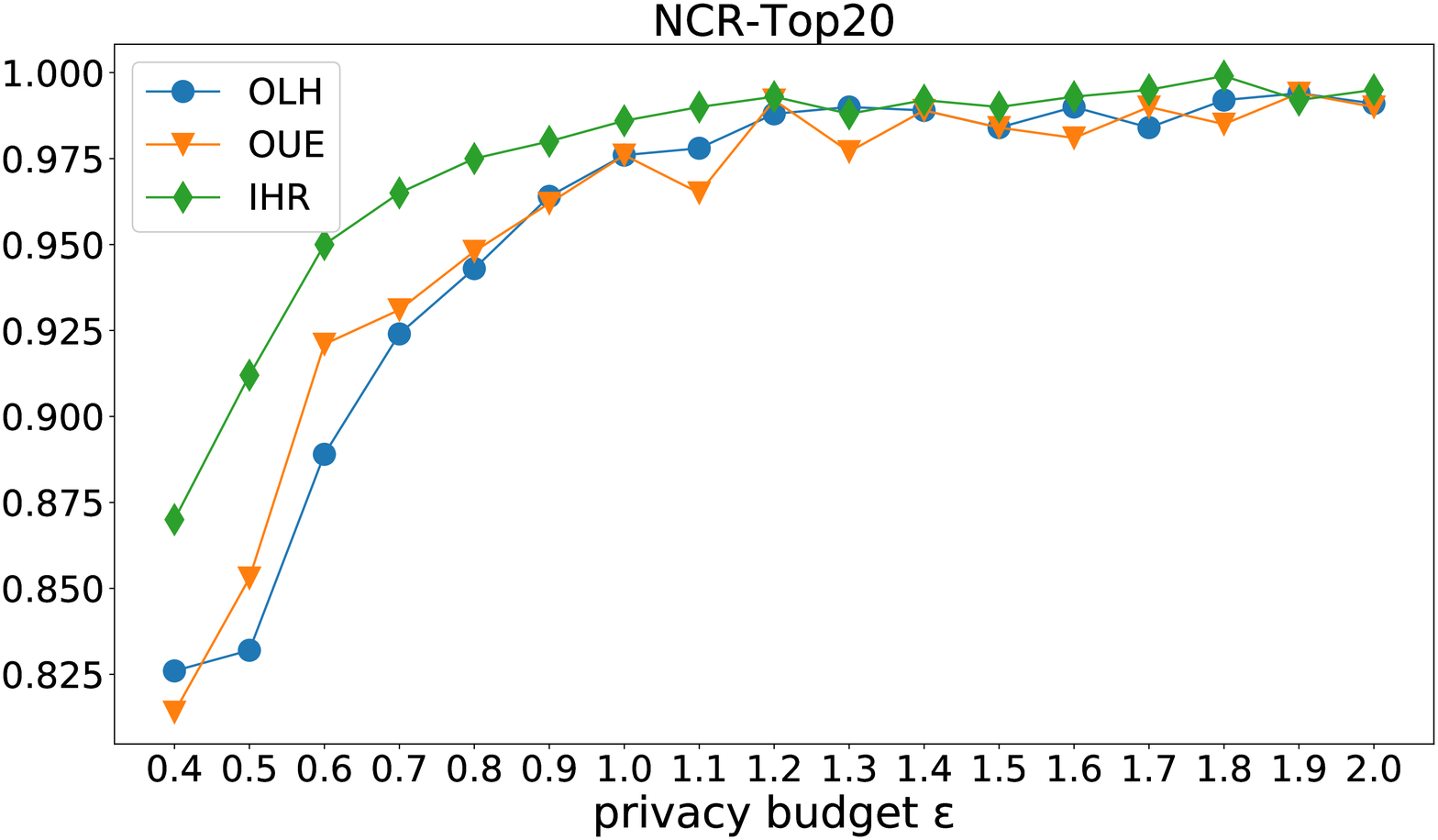}
		\label{fig:Online_Top20_NCR}}
	\subfigure[Cohort for Top 20]{
		\includegraphics[width=0.29\textwidth]{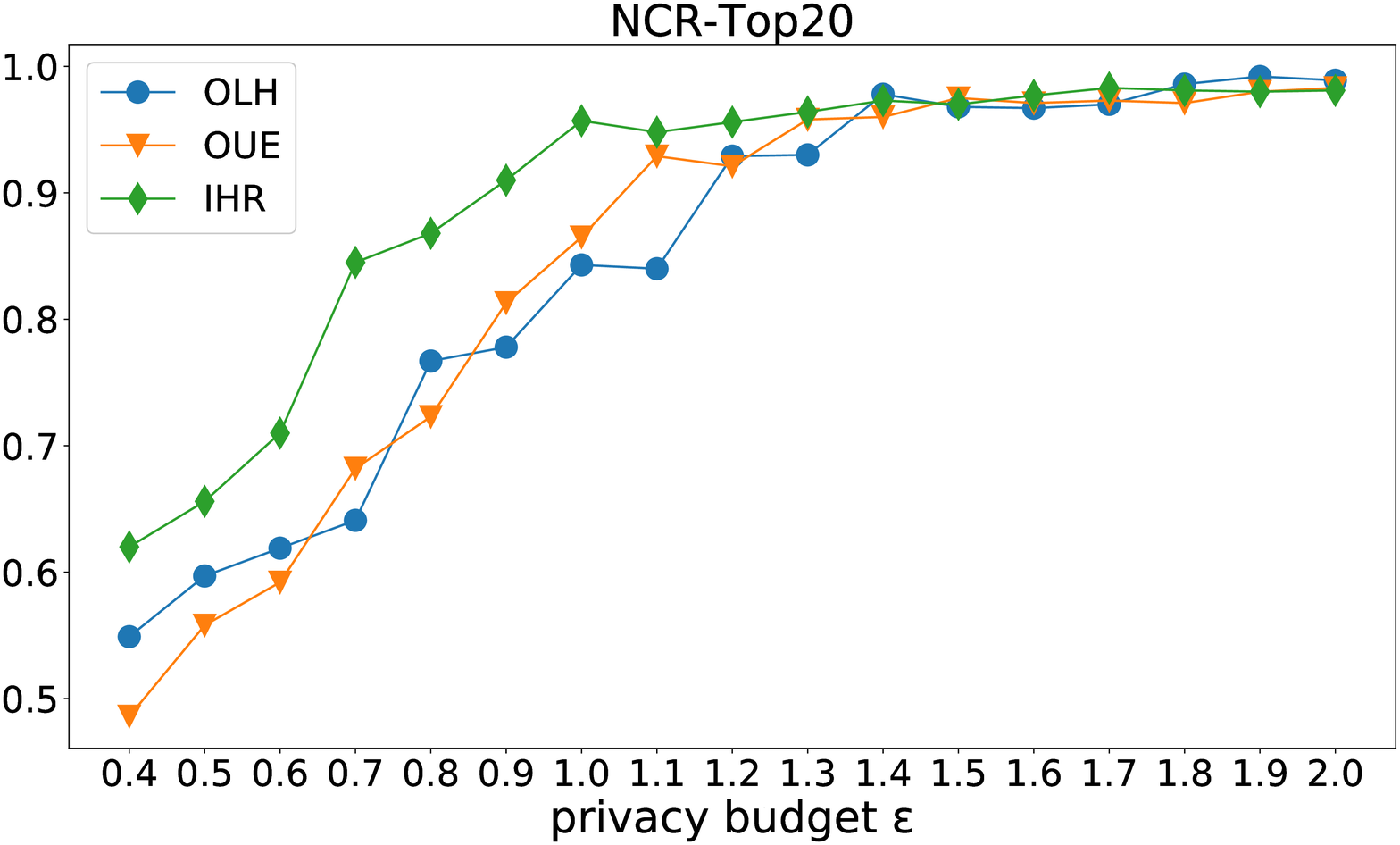}
		\label{fig:Cohort_Top20_NCR}}
	\subfigure[ZipfData for Top 50]{
		\includegraphics[width=0.29\textwidth]{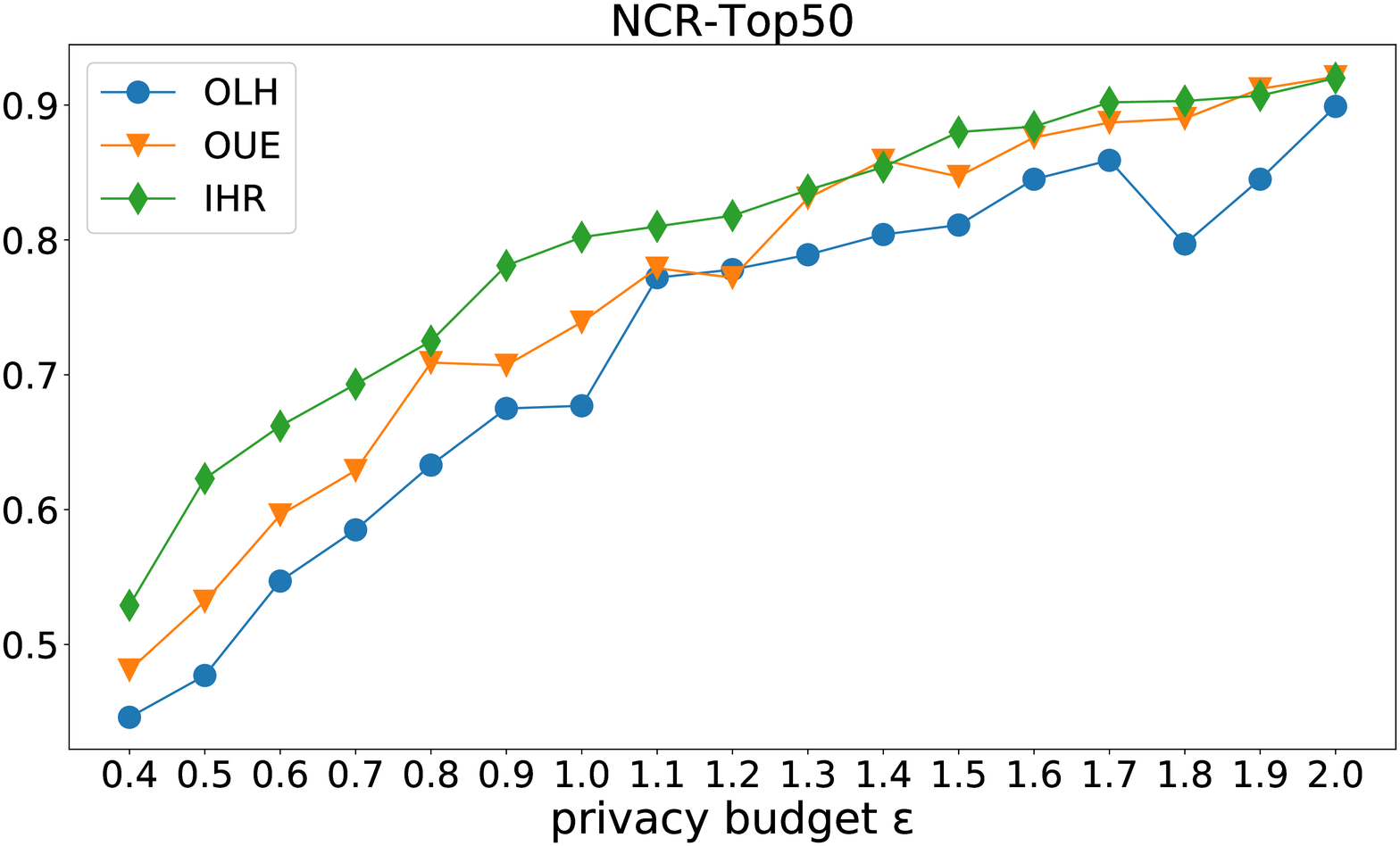}
		\label{fig:ZipfData_Top50_NCR}}
	\subfigure[Online for Top 50]{
		\includegraphics[width=0.29\textwidth]{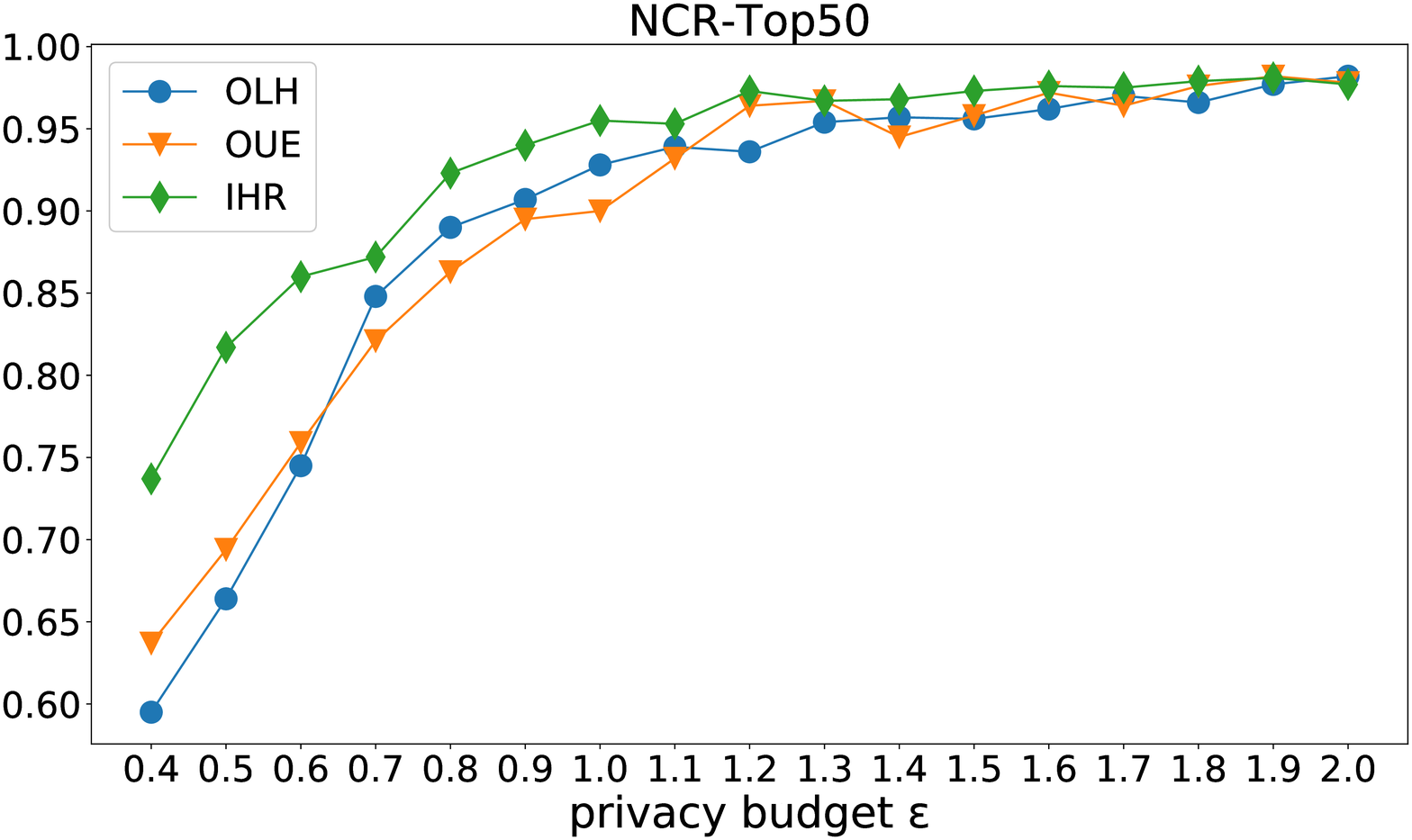}
		\label{fig:Online_Top50_NCR}}
	\subfigure[Cohort for Top 50]{
		\includegraphics[width=0.29\textwidth]{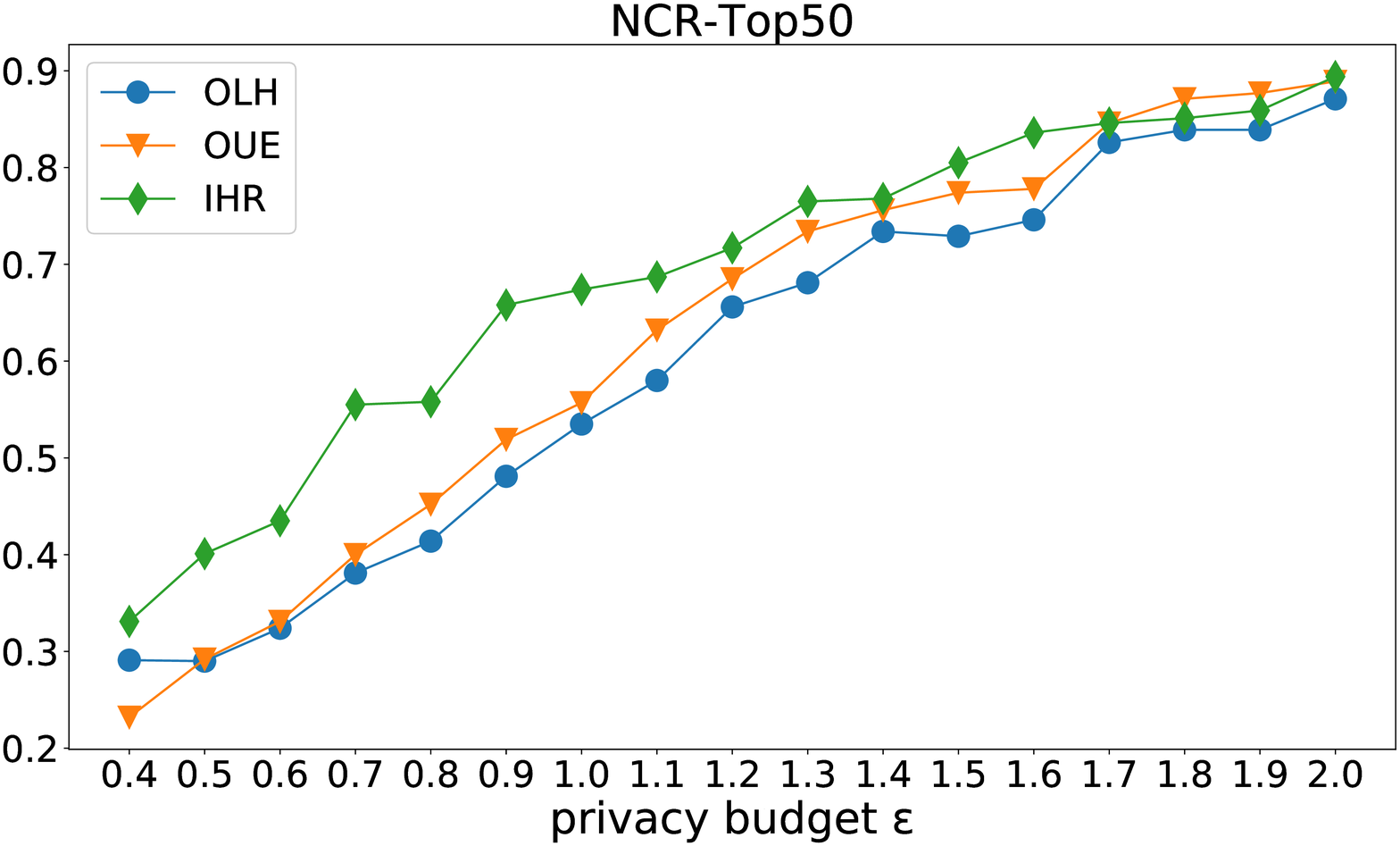}
		\label{fig:Cohort_Top50_NCR}}
	\subfigure[ZipfData for Top 100]{
		\includegraphics[width=0.29\textwidth]{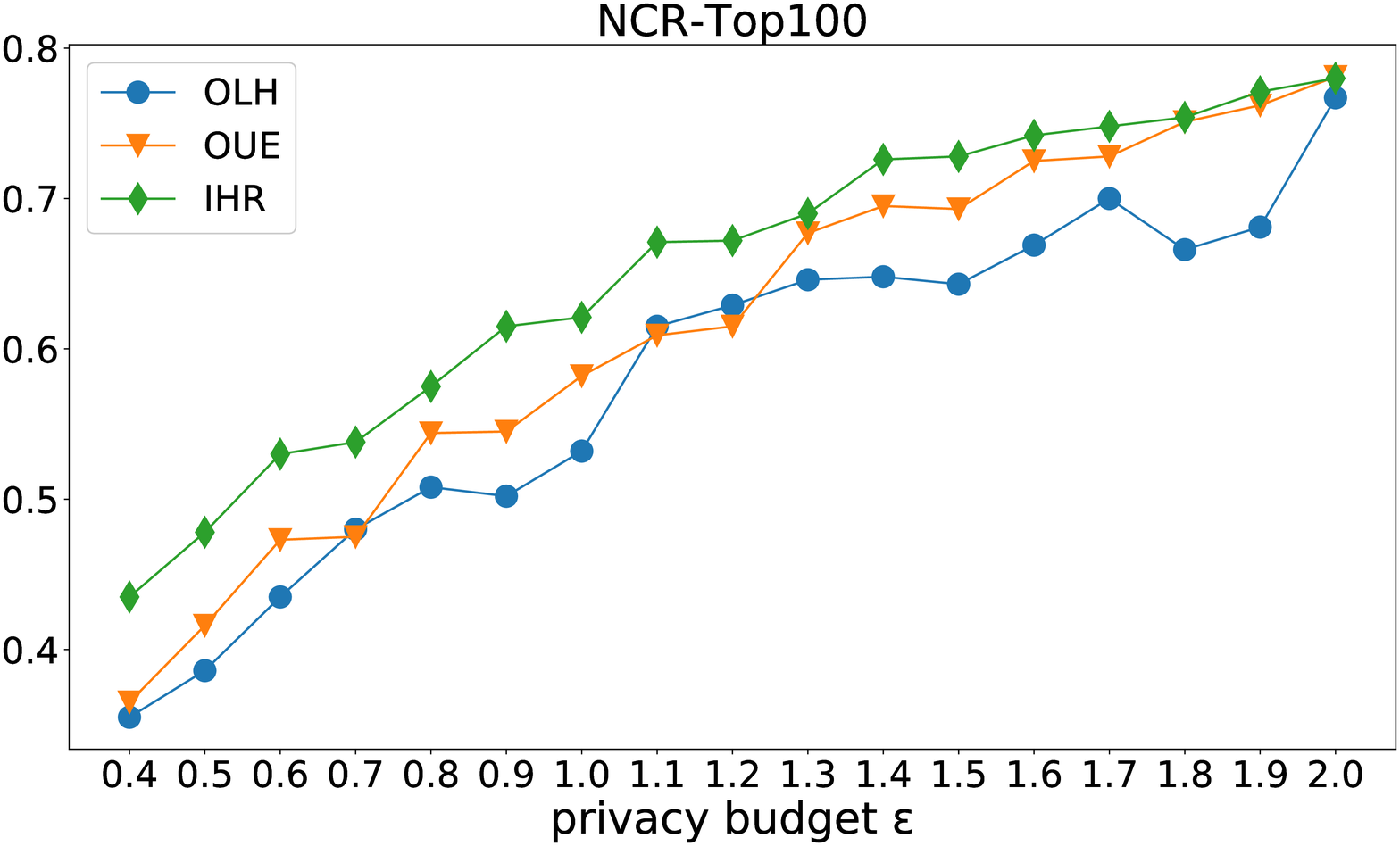}
		\label{fig:ZipfData_Top100_NCR}}
	\subfigure[Online for Top 100]{
		\includegraphics[width=0.29\textwidth]{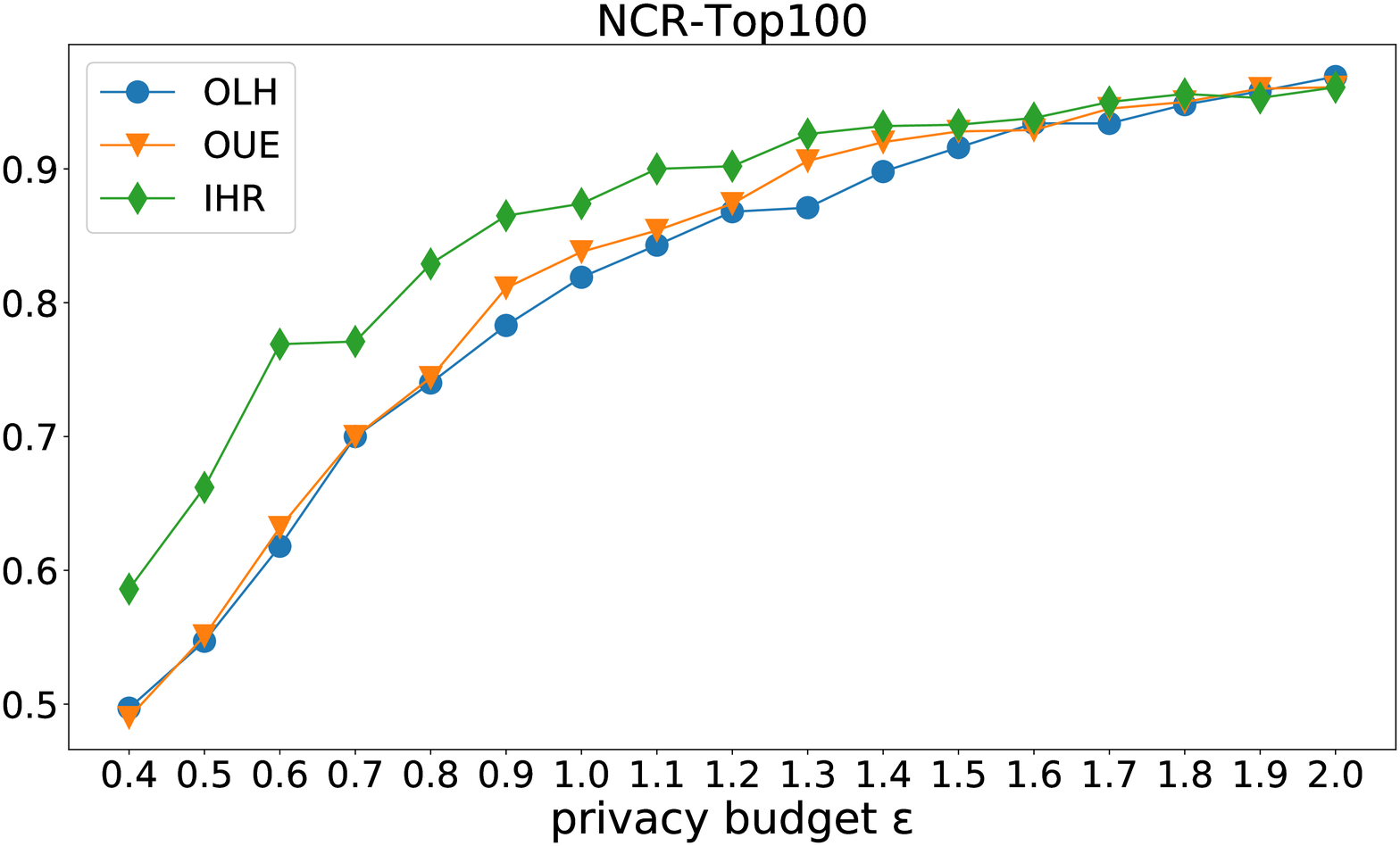}
		\label{fig:Online_Top100_NCR}}
	\subfigure[Cohort for Top 100]{
		\includegraphics[width=0.29\textwidth]{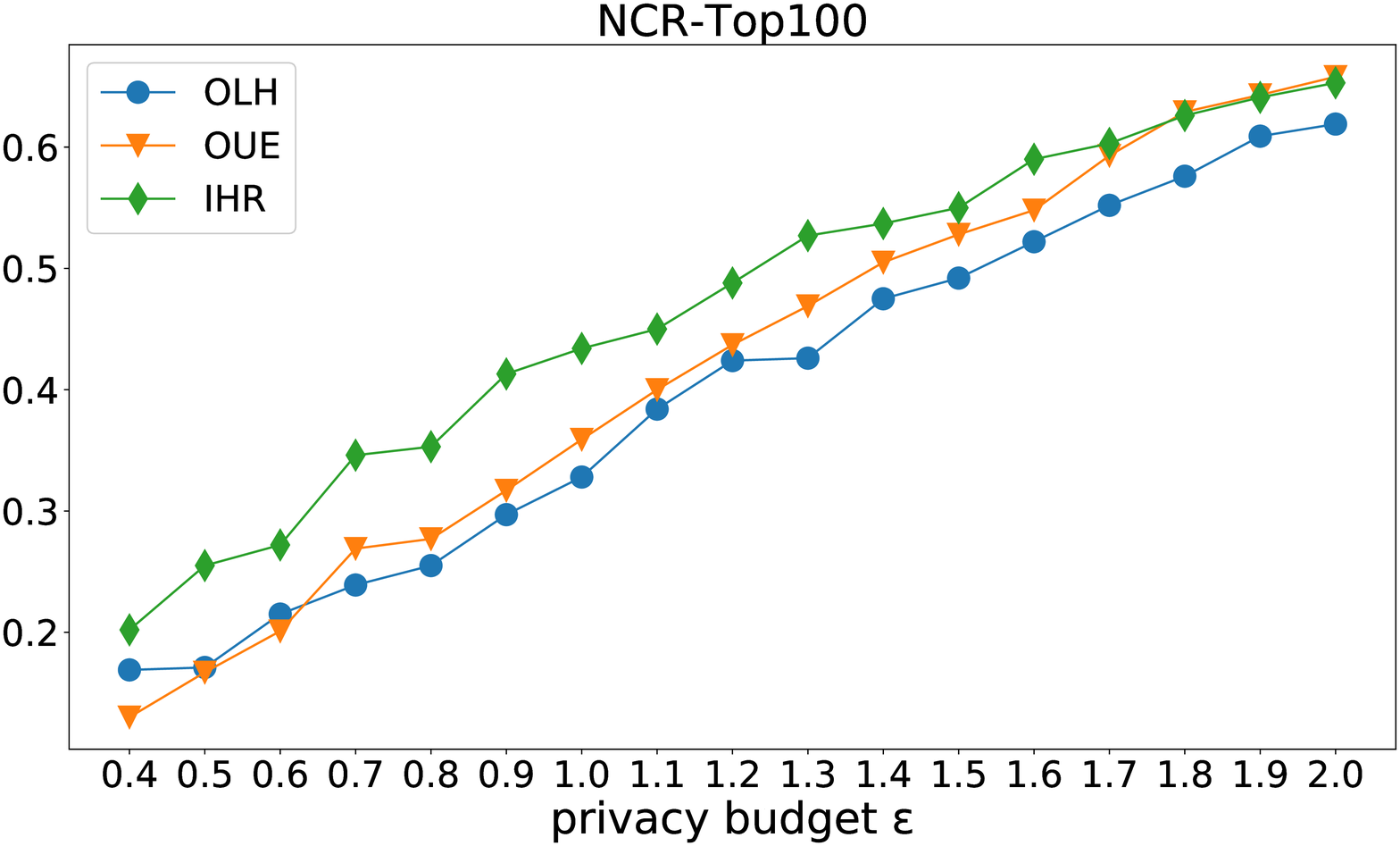}
		\label{fig:Cohort_Top100_NCR}}
	
	\caption{NCR for each dataset}
	\label{Fig: result_NCR}
\end{figure*}

Third, Figure \ref{Fig: result_RE} presents the related error (RE) for the FO for all datasets. In particular, the first row in Figures \ref{fig:ZipfData_Top20_NCR}, \ref{fig:Online_Top20_NCR}, and \ref{fig:Cohort_Top20_NCR}, the second row in Figures \ref{fig:ZipfData_Top50_NCR}, \ref{fig:Online_Top50_NCR}, and \ref{fig:Cohort_Top50_NCR}, and the third row in Figures \ref{fig:ZipfData_Top100_NCR}, \ref{fig:Online_Top100_NCR}, and \ref{fig:Cohort_Top100_NCR} show trendlines with increasing privacy budgets when the top 20, top 50, and top 100 values are selected, respectively.  The trendlines in the top 20 and top 50 figures decrease as the privacy budget $\varepsilon$ increases, and for these three datasets, although the trendlines in the top 100 figures show drastic fluctuations, the FHR mechanism trendline (green) is the lowest in most cases. According to the definition of RE, the lower the trendline, the more accurate the FO. Thus, the mechanism FHO performs best with respect to RE.

Fourth, Figure \ref{Fig: result_NCR} presents the NCR for the FO for all datasets. In particular, the first row in Figures \ref{fig:ZipfData_Top20_NCR}, \ref{fig:Online_Top20_NCR}, and \ref{fig:Cohort_Top20_NCR}, the second row in Figure \ref{fig:ZipfData_Top50_NCR}, \ref{fig:Online_Top50_NCR}, and \ref{fig:Cohort_Top50_NCR}, and the third row in Figures \ref{fig:ZipfData_Top100_NCR}, \ref{fig:Online_Top100_NCR}, and \ref{fig:Cohort_Top100_NCR} show the trendlines with increasing privacy budgets when the top 20, top 50, and top 100 values are selected, respectively.  The trendlines in these figures increase as the privacy budget $\varepsilon$ increases, and the trendlines for FHR (green) are the highest. According to the definition of NCR, the higher the trendline, the more high-frequency items the FO collects. Thus, mechanism FHO performs best with respect to NCR.

\textbf{Summary.} The experimental results demonstrate that FHR performs well in the case of a small privacy budget. The reason for the above observations is FHR’s effective and efficient transmission strategy based on sampling. Random sampling of the Hadamard matrix is used to make transmission more efficient when $\varepsilon$ is small; however, the real frequency cannot be obtained when $\varepsilon$ is very large. This illustrates the advantage and disadvantage of sampling. Meanwhile, this study provides an improved mechanism, FHR, while satisfying SFLDP. There remains substantial room for improvement of FLDP and SFLDP in future work.

\section{Conclusion}

LDP is overly conservative, preventing its application in practice. In this study, we proposed FLDP and SFLDP to weaken LDP in order to provide more flexible notions. LDP requires considerable noise in order to hide one input among all inputs. Under our notion, we classify the input based on the output in order to obtain more accurate statistics without decreasing disturbance parameters while ensuring local data privacy. Based on practical applications, this study presented the new FHR mechanism, which considers privacy protection, communication, and computational complexity. Finally, theoretical analysis and experiments demonstrated that FHR can obtain an accurate FO. 
\section{Acknowledge} 
This work is supported by National Key RD Program of China (No.2018YFB1004401) and NSFC under the grant No. 61532021, 61772537, 61772536, 61702522).



\end{document}